\documentclass[12pt,preprint]{aastex}
\usepackage{emulateapj5}
\shorttitle{Planet Frequencies in $\omega$ Centauri}
\shortauthors{Weldrake, Sackett \& Bridges}

\begin{document}
\title{The Frequency of Large Radius Hot and Very Hot Jupiters in $\omega$ Centauri}

\author{David T F Weldrake} 
\affil{Max Planck Instit\"ut f\"ur Astronomie, K\"onigstuhl 17, D-69117, Heidelberg, Germany}
\email{weldrake@mpia-hd.mpg.de}

\author{Penny D Sackett}
\affil{Research School of Astronomy and Astrophysics, The Australian National University, Mount Stromlo Observatory, Cotter Road, Weston Creek, ACT 2611, Australia}
\email{penny.sackett@anu.edu.au}

\author{Terry J Bridges}
\affil{Physics Department, Queen's University, Kingston, Ontario, Canada K7L 3N6}
\email{tjb@astro.queensu.ca}

\begin{abstract}
We present the results of a deep, wide-field search for transiting `Hot Jupiter (HJ)' planets in the globular cluster $\omega$ Centauri. As a result of a 25-night observing run with the ANU 40-inch telescope at Siding Spring Observatory, a total of 109,726 stellar time series composed of 787 independent data points were produced with differential photometry in a 52$'$$\times$52$'$ (0.75 deg$^2$) field centered on the cluster core, but extending well beyond. 
Taking into account the size of transit signals as a function of stellar radius, 45,406 stars have suitable photometric accuracy ($\le$0.045 mag to V$=$19.5) to search for transits. Of this sample, 31,000 stars are expected to be main sequence cluster members. All stars, both cluster and foreground, were subjected to a rigorous search for transit signatures; none were found. Extensive Monte Carlo simulations based on our actual data set allows us to determine the sensitivity of our survey to planets with radii $\sim$1.5R$\rm{_{Jup}}$, and thus place statistical upper limits on their 
occurrence frequency $F$.  Smaller planets are undetectable in our data. At 95$\%$ confidence, the frequency of Very Hot Jupiters (VHJs) with periods P satisfying 1d~$<$~P~$<$~3d can be no more than $F_{\rm VHJ} < 1/1040$ in $\omega$ Cen. For HJ and VHJ distributed uniformly over the orbital period range 1d~$<$~P~$<$5d, $F_{\rm VHJ+HJ} < 1/600$. Our limits on large, short-period planets are comparable to those recently reported for other Galactic fields, despite being derived with less telescope time.
\end{abstract}

\keywords{globular clusters: individual (NGC 5139, $\omega$ Centauri) - planetary systems - techniques: photometric}

\section{Introduction}
The identification and subsequent study of extrasolar planets has become a subject of intense interest in recent years. To date, $\sim$250 giant planets have been discovered\footnote{http://exoplanet.eu/}, ranging in mass from Neptune to greater than Jupiter. These new worlds are altering our understanding of the formation and evolutionary processes of giant planets in the immediate solar neighborhood. Currently, the majority of them have been found using radial velocity (RV) techniques, which favours the detection of close-in, massive planets.
 
Statistical analysis of current RV detections indicates that 1.2$\%$$\pm$0.3$\%$ of nearby F,G and K stars are orbited by `Hot Jupiter' (HJ) planets, those with orbital periods of only a few days ($\le$0.1 AU) and minimum masses approximately equal to that of Jupiter \citep{M2005}. Such discoveries have challenged traditional ideas of planetary evolution, implying that a rapid migration of the planet takes place soon after formation. 

Planet frequency from radial velocities appear to depend on the metallicity of the host star \citep{G1997,L2000,S2001,FV2005}. However, there is very little observational evidence for a lower planetary frequency at quite low metallicities, due to the bias of radial velocity detections in association with bright nearby high metallicity stars. Hence dedicated surveys in low metallicity environments, like globular clusters, can provide information to help understand this relationship in a more robust manner. Indeed, does low metallicity halt planet formation or just affect the planetary migration process? 

Of the currently known planets, $\sim$90$\%$ have only a minimum mass assigned to them, due to the unknown inclination of the planetary orbit, and unknown radii and densities. These quantities can be measured for transiting planets.
Due to its short orbital period, each HJ has a non-negligible probability of transiting its host star that depends on the orbital separation and the ratio between the stellar and planetary radii. Typical transit depths and durations are $\sim$1.5$\%$ and $\sim$2 hours respectively. With precise photometry these transiting systems can be identified in the field, leading to direct measurements of the planetary radius (provided the stellar radius is known) from the depth of the transit dip. Studies of the planetary atmosphere may be attempted if transits occur, and when coupled with RV measurements, accurate mass and density determinations can be made.

Currently, 28 transiting exoplanets are known \citep{C2000,He2000,K2003,K2005,B2004,B2005,A2004,P2004,S2005,MCC2006,Bakos2006,Bakos2007,Bakos2007c,OD2006,OD2007,CC2006,Sahu2006,Bu2007,G2007,Mand07,Barb2007,Kov2007,N2007}, only a handful of which were first discovered through radial velocity searches. Despite an increasing number, the detection rate of planets from transit searches is significantly lower than initially expected eg, \citet{H2003}. This lack of detections is due, in part, to observing strategy: a long observing window ($\ge$1 month equivalent) with a dedicated telescope coupled with a wide field and high temporal resolution are needed to sample enough stars frequently enough to allow the detection of a transit. Also the production of a large enough number ($\sim10^{5}\rightarrow10^{6}$) of high-quality ($\le$0.02 mag) lightcurves of dwarf stars coupled with a sensitive detection algorithm with low false alarm rates is non-trivial.

Recently, \citet{Gouldetal2006} analyzed OGLE~III transit surveys in Galactic 
fields and concluded that the occurrence frequencies of the detected planets 
in these surveys is not statistically different from that found in 
RV surveys of nearby stars. However, they did conclude that the frequency of 
HJ planets with periods P satisfying 3$<$P$<$5 days ($F =$~1/320(1$^{+1.37}_{-0.59}$) at a 90$\%$ confidence) was statistically different from and larger than that of Very Hot Jupiter (VHJ) planets (1$<$P$<$3 days), which have $F = $1/710(1$^{+1.10}_{-0.54}$) at 90$\%$ confidence.  Since the OGLE~III surveys detected no planets with radius larger than 1.3R$_{\rm{Jup}}$,they placed upper limits on the occurrence frequency of larger worlds.

Transit searches, unlike RV, are not limited to the immediate solar neighborhood and can be used to measure relative planet frequencies in various regions of the Galaxy, providing information on the role environmental effects play on HJ planet formation. Early predictions for the success of transit searches in open clusters was presented by \citet{J1996}, indicating that with a large amount of telescope time planets could be detected via the transit technique in nearby clusters. More recently, \citet{PG2005} presented an analysis of the prospective harvest of cluster transit surveys by discussing the observational techniques and methods to maximise the chances of a detection. They concluded that due to their mass functions, the most populous, nearby and bright clusters have the greatest chance of yielding a planet. \citet{PG2006} then went on to discuss specifically the detection of short period `Hot Earth' and `Hot Neptune' planets. The detection yields for various nearby clusters with various instruments was estimated. \citet{Ai2007} agree that small-aperture wide field surveys targetting nearby clusters have the potential to discover transiting Hot Neptune planets.

Transit searches have been undertaken in bright metal-rich open clusters including STEPPS \citep{Bu2005}, UStAPS \citep{S2003,Ho2005}, PISCES \citep{M2003,Mo2005,Moch2006}, $\it{Monitor}$ \citep{A2007}, EXPLORE-OC \citep{V2005} and in the Praesepe cluster (M44) with KELT \citep{Pepper07}. Searches have also been performed in the general Galactic field \citep{U2002,U2004,H2005,Ka2005,W2007a} and toward the Galactic Bulge \citep{Sahu2006}. If cluster candidates are confirmed as planetary in nature, difficult if fainter than V$\sim$17.0, they can provide information on the timescales of HJ formation and subsequent migration. Null results of high significance allow planet frequency upper limits to be estimated.

Globular clusters provide an excellent opportunity to study the effects of environment on planetary frequency. Two bright, nearby southern clusters, 47 Tuc and $\omega$ Centauri, have stars in sufficient numbers ($\sim$10$^{5}$) and brightness (V$\leqslant$17) for meaningful statistics to be gained using ground-based telescopes of moderate aperture. 47 Tuc was previously sampled for planetary transits, resulting in a high significance ($>$3$\sigma$) null result in both the cluster core \citep{G2000} and in the outer halo \citep{W2005}. These two results strongly indicate that system metallicity - not crowding - is the dominant factor determining HJ frequencies in this cluster.

This paper presents the results of a dedicated transit search in the second cluster, $\omega$ Centauri, in order to test further the dependence of planetary frequency on stellar metallicity and crowding. Omega Centauri has only 1/10$^{\rm{th}}$ the core density of 47 Tuc yet contains five times the total mass (5.1$\times$10$^6$M$_{\odot}$, \citet{Meylan1995}). Due to its low stellar density compared to other globular clusters and long stellar interaction timescale, a null result for $\omega$ Centauri can be used to test the relative importance of stellar metallicity over density in the formation of giant planets.

Omega Centauri ($\omega$ Cen, NGC 5139) has been subjected to intense research over the years. The cluster is unique among globular clusters in that it displays a distinct spread of metallicity among its stars \citep{DW1967,NB1975,Lee1999,P2000,Sol2005}, due to an extended period of star formation and chemical enrichment. Using He abundances, \citet{N2004} has shown that the cluster has three distinct stellar populations, with metallicities of $-$1.7, $-$1.2 and $-$0.6 dex, corresponding to 0.80, 0.15 and 0.05 of the total population respectively.

The cluster has a highly-bound, retrograde orbit \citep{D1999} and is by far the most massive of the globular clusters \citep{Meylan1995}. Indeed, these vagaries have led to the theory that the cluster had an external origin, being the left-over remains of a tidally disrupted dwarf galaxy \citep{BF2003,IM2004,BN2005}. With its relative proximity, $\omega$ Cen presents a statistically significant number of upper main sequence stars that can be searched for transiting HJ planets. 

Here we present the result from a vigorous search for the transit signatures of large planets on 45,406 lightcurves in a 0.75deg$^2$ field centered on $\omega$ Cen. The same set of observations yielded a total of 187 variable stars in the field, 81 of which are new discoveries, and are presented in a companion paper \citep{W2006}. Furthermore, we observed a control field in the Lupus Galactic Plane to test the data reduction and transit identification strategies. Analysis for this field is ongoing, but has led to the identification of several transit candidates, of which none similar were seen in the $\omega$ Cen dataset. One candidate in particular has excellent prospects for being a new Hot Jupiter planet \citep{W2007a}.

Section 2 of this paper describes our observational strategy and data reduction details. Section 3 details how the photometry was obtained for both the crowded core regions and outer halo parts of the dataset. The cluster Color Magnitude Diagram dataset (along with astrometry) are also briefly discussed. Section 4 describes the stellar parameters of the cluster stars that were searched for transits, and the expected characteristics of the transits themselves. The total number of stars in the field (both in the cluster and the foreground galactic disk) is also calculated. Section 5 describes the transit detection algorithms used in our search and our removal of systematics in the photometry. Our Monte Carlo simulations to derive expected recovery of real transits and false alarms are described in Section 6, and their application to estimate our HJ sensitivity outlined in Section 7. The results of our transit search in $\omega$ Cen are presented in Section 8, with discussion, 
comparison to the literature and interpretation in Section 9. We conclude in Section 10.

\section{Observations and Data Reduction}
Our dataset was produced using the Australian National University (ANU) 40-inch (1m) telescope located at Siding Spring Observatory, fitted with the ANU Wide Field Imager (WFI). This telescope and detector combination permits a 52$'$$\times$52$'$ (0.75 deg$^2$) field of view, which was centered on the cluster core. 

WFI is capable of sampling a large fraction of the cluster in a single exposure, as the field extends to 50$\%$ the cluster tidal radius \citep{Harris96}, allowing a large time series dataset to be produced with only a single pointing. The WFI detector consists of a 4$\times$2 array of 2048$\times$4096 pixel back-illuminated CCDs in an 8K$\times$8K arrangement. The pixel scale at the telescope Cassegrain focus is 0$''$.38 pixel$^{-1}$, permitting suitable sampling of the point spread function (PSF).

In order to increase the likelihood of transit detection, an observing strategy was employed to maximize the temporal resolution of the dataset while keeping the resultant signal-to-noise ratio (S/N), and hence photometric precision, high. In other words, deep images with short exposure times are necessary. In order to do this with a 1m telescope, a special broad-band filter was constructed to cover the combined wavelength range of Cousins V and R. 

A five minute exposure of a V$=$18.5 star (typical of a main sequence star in the cluster) with this filter has a photon noise S/N of 220 in 7-day moon and 2$''$ seeing. In order for a typical HJ (transit depth $\sim$0.015 mag) to be detected at the 3$\sigma$ level, an effective S/N of 200 is required, which is thus obtainable with this V+R filter on timescales that would well resolve the expected $\sim$2-hour transits. This same telescope and detector combination was used successfully for a transit search in 47 Tucanae, which led to a high significance null result in that cluster \citep{W2005}.   

The globular cluster $\omega$ Centauri was observed for 25 contiguous nights, from 2003 May 2 to 2003 May 27 with a field center (J2000.0) of $\alpha$ $=$ 13$^{h}$26$^{m}$45.89$^{s}$, $\delta$ $=$ $-$47$^{\circ}$28$'$36.7$''$ (J2000.0). An exposure time of 300s was used throughout the observing run with the V+R filter. Each image was checked individually for quality after readout at the telescope. If an image displayed bad seeing ($\ge$3$''$), bad focus or cosmetic problems (such as satellite trails, intermittent clouds or other adverse effects) it was discarded. After this quality control, a total of 875 images were obtained over the 25-night run, the dataset having an average temporal resolution of six minutes and covering nine hours on a typical night.

Initial image reduction was performed using standard reduction practices with the MSCRED package of IRAF\footnote{IRAF is distributed by National Optical Observatories, which is operated by the Association of Universities for Research in Astronomy, Inc., under cooperative agreement with the National Science Foundation.}. This procedure included region trimming, overscan correction, bias correction, flat-fielding and dark current subtraction. The reduced images were then checked for quality before entering the photometric pipeline. Of the 875 images obtained, 90$\%$ (787 images) were deemed suitable for use in the production of the time series dataset, as indicated by their small telescope offsets and good seeing. 

\section{Photometry and Photometric Accuracy}
High precision photometry can be obtained on faint targets in the crowded field of $\omega$ Cen by performing differential photometry. This method was originally described as an optimal Point-Spread-Function (PSF) matching algorithm by \citet{AL98}, and was subsequently modified by \citet{Woz2000} for use in detecting microlensing events. A detailed description of the Difference Image Analysis (DIA) method and software pipeline can be found in Wozniak's paper; only the main steps are summarized here. 

The process of matching the stellar PSF in a large database of images reduces dramatically the systematic effects due to varying atmospheric conditions on the photometric precision, allowing the detection of small brightness variations in faint targets with ground-based observations. DIA is also an excellent photometric method for dealing with crowded fields. Since a larger number of pixels contain information on any PSF differences as the number of stars increases, there is an improvement in the PSF-matching process. Baseline flux measurements of the stars are made via profile fitting on a master template frame, which is produced by median-combining a large number (40$+$ in our case) of the best quality images with small offsets. This flux measurement is used as the zero-point in the photometric time series for an individual star. 

Stellar positions are determined from a reference image, usually the image with the best seeing conditions; and all other images in the dataset (including the template) are shifted to match. The best PSF-matching kernel is then found for each image, and each registered image is subtracted from the template. The residual subtraction image is generally dominated by photon noise. Any object that has changed in brightness between the image and the template is given away as a bright or dark spot. 

The pixel coordinates of all visible stars were determined separately on the reference frame via DAOFIND within IRAF, and the profile photometry was then extracted from the subtracted frames at those determined positions. 

Differential photometry produces a time series measured in differential counts, a linear flux unit from which a constant reference flux (taken from the template) has been subtracted. To convert this to a standard magnitude system, the total number of counts for each star was measured using the PSF photometry package of DAOPHOT within IRAF, with the same images and parameters as used in the photometry code. The time series fluxes were then converted into magnitude units via the relation:

\begin{center}
$\Delta m_i = -2.5 \log [(N_i+ N_{\rm ref,\it{i}}) / N_{\rm ref,\it{i}}]$
\end{center}

\noindent where $N_{\rm ref,\it{i}}$ is the total flux of star $\it{i}$ on the template image and $N_i$ is the difference flux in the time series as produced with the photometric code. 

When combining differential fluxes with DAOPHOT-derived photometry on the reference image, it is important to correct for errors based on the individual apertures used. The scaling between the two fluxes was determined via performing an aperture correction on the DAOPHOT magnitudes for the cluster stars, as per the method described in Appendix B of \citet{H2004}. We found that our PSF magnitudes were consistently 0.09 magnitudes brighter than the aperture-derived values (using the same aperture values as in the differential photometry). We corrected for this by shifting our magnitude zero-point to 25.0$-$ 0.09 $=$ 24.91.

A total of 109,726 stellar photometric time series were produced across the whole WFI field, each containing 787 independent data points, which then became the subject of analysis.

\subsection{Photometry of the cluster core}
Data from each of the four outer CCDs of WFI were divided in half for DIA analysis, with each half producing an average of 9,500 time series. For the crowded core of the cluster, where the number of stars becomes very large, computational limitations necessitated a different strategy. For the core regions, the images were analysed in 360 individual subframes, 90 per CCD. The locations of these subframes were chosen so that no stars would be lost at the edges of each individual subframe, and the entire core region of the cluster was covered. Data of sufficient quality could not be obtained in regions affected by telescope offsets during observing (typically a 160 pixel border surrounding each CCD. The SYSREM systematics removal package of \citet{T2005} was applied to all resulting lightcurves. 

Fig.\space\ref{rmsplot} presents the resultant DIA-derived photometric precision, measured as root-mean-square scatter ($\it{rms}$), for the 97,935 stars that were cross-identified with the cluster CMD dataset. The position of the cluster main sequence turnoff (MSTO) is marked to indicate where the cluster stars become members of the main sequence. Objects to the left of this line are likely Red Giant Branch and foreground Galactic Disk stars. The photometric uncertainty is 1$\%$ at V$=$17.4, rising to 4$\%$ at V$=$19.0. 

Also overlaid on Fig.\space\ref{rmsplot} is the expected depth of a transit for a 1.5R$_{\rm{Jup}}$ planet as a function of V magnitude of the parent star. The total star, background and residual noise contribution is also plotted (thin locus of red points). The photometry is photon-noise-limited for V$\leqslant$17.0 and sky+residuals dominated to fainter magnitudes. We define our lower limit as lying at the magnitude value where all of the stars in our dataset have photometric uncertainty larger than the expected transit depth. It can be seen on Fig.\space\ref{rmsplot} that this limit is reached at V$=$19.5.

\subsection{Color Magnitude Diagram and Astrometry}
Using the same telescope/detector combination, a V, V-I color magnitude diagram (CMD) totalling 203,892 stars was produced for the observed field, enabling the detected transiting systems to be placed on the standard V and I magnitude system. This aids in determining their likely nature. The total CMD dataset is presented in our variable star companion paper \citep{W2006}. Fig.\space\ref{transcmd} presents the diagram for the upper main sequence region that was targeted for the transit search. The photometric errors reported by DAOPHOT in both V and V-I are marked as errorbars as a function of V magnitude. 

As standard field data was of unacceptable quality, the CMD was calibrated by matching stellar astrometry from our catalog to that of \citet{C2004} (also taken with the ANU 40-inch and WFI combination in V and I). The difference in V and I between our uncalibrated data and the \citet{C2004} calibrated data was measured for each of the matched stars (totalling more than 20,000) in each CCD independently. The resulting calibration accuracy is $\le$0.03 magnitudes. Variable stars found in the monitoring data were identified on the CMD by visually identifying the star on the template image and comparing it to the V-band CMD image for the respective CCD. 

Astrometry was obtained for a total of 212,959 stars identified in the V band image of the cluster and 243,466 stars in the I-band. A search of the USNO CCD Astrograph Catalog (UCAC1) was carried out for astrometric standard stars within the field. Several hundred such stars were successfully identified, producing an accurate determination of the astrometric solution for the stars in each CCD independently; the resulting calibration accuracy was 0.25$''$. The extent of our single WFI field of view can be seen in Fig.\space\ref{chips}, plotted as $\Delta$RA and $\Delta$Dec in degrees from the position of the cluster core. For comparison, the locations of the cluster core radius (inner ellipse), half-mass radius (middle ellipse), and the location of 50$\%$ of the cluster tidal radius are also marked overlying the location of the eight WFI CCDs.

\section{Stellar Parameters, Transit Expectations and Total Number of Stars}
In order to determine the expected depths and durations of planetary transits (which are necessary for the Monte Carlo simulations) against main sequence cluster stars, the radii of those stars were first calculated as a function of magnitude. This was done by producing three theoretical \citet{Y2003} isochrones, each with metallicity and age values for $\omega$ Cen as taken from \citet{N2004} to simulate the cluster stellar content. The majority of the stars ($\sim$80$\%$) are expected to lie on the most metal-poor ([Fe/H]$=$$-$1.7) sequence, which was used for our analysis. These three isochrones and the corresponding parameters can be seen superimposed on Fig.\space\ref{transcmd}.

Table 1 presents the resulting stellar parameters as determined from the isochrone for this metal poor population for stars with V$=$17.0$\rightarrow$20.0, namely those stars on the cluster turnoff and upper main sequence. Tabulated are the apparent magnitude, the stellar mass, temperature and luminosity (in Solar units), the stellar surface gravity and the corresponding absolute magnitude. The stellar radius has been calculated from the surface gravity values and is also tabulated. The stellar radius increases rapidly for cluster stars brighter than the main sequence turnoff (at V$\sim$17.5), hence transit visibility becomes reduced for the brightest stars.

From these calculated radii, the expected depths in magnitude units (Dep) and durations in hours (Dur) for a transiting planet were found via:

\vspace{-0.8cm}
\begin{center}
\begin{equation}
\rm{Dep} \sim \left( \frac{R_{p}}{R_{\ast}} \right) ^2
\end{equation}
\end{center}
\noindent where $R_{p}$ is the radius of the planet and $R_{\ast}$ is the radius of the star, and:

\vspace{-0.6cm}
\begin{center}
\begin{equation}
\rm{Dur} = 1.412M_{\ast}^{-1/3}R_{\ast}P_{orb}^{1/3}
\end{equation}
\end{center}
\noindent where $M_{\ast}$ and $R_{\ast}$ are the stellar mass and radius in solar units, and $P_{orb}$ is the orbital period of the planet in days, taken from \citet{G2000}. We assume a period of 3.3d, typical of HJ planets found in the solar neighborhood. The duration value incorporates a $\pi$/4 reduction in transit duration, corresponding to the average chord length across a stellar disk. A centrally-crossing transit would have a duration 1.28 times longer than the values determined here. This relation can be used to infer the planetary radius of any detected candidates with a known period and total duration. The orbital inclination is also determined in that calculation.

The transit depth and duration values as calculated via the Yi isochrones are displayed for planets of 1.5$\rm{R_{Jup}}$ in Fig.\space\ref{transcmd} for various V magnitudes of the cluster main sequence. Also overplotted is the location in brightness at which various photometric precisions can be expected. The precision becomes comparable to the expected depth of a 1.5$\rm{R_{Jup}}$ planet at V$\sim$19.5.

The expected transit depths are superimposed on Fig.\space\ref{rmsplot} for all cluster stars. The line indicates 1.5$\rm{R_{Jup}}$ planet depths and we do not have the photometric accuracy to search for planets smaller than this limit. The transit depth becomes very small once the stars leave the cluster turnoff (hence becoming physically larger) and rapidly become undetectable. For larger planets, the transit depth is similar to the photometric uncertainty, as stars are smaller further down the main sequence. The optimal zone for transit detection is V$=$17.0$\rightarrow$19.0, with the limit at V$\sim$19.5, where photometric errors dominate the lightcurve and the transit depth is greater than the photometric uncertainty. We do not consider the signal-to-noise of the transits in our limit determination, only the magnitude limit for which the per-data point photometric error becomes larger than the expected transit depth. 

With this knowledge, the total number of stars available for analysis was determined for this planetary radius limit. For 1.5$\rm{R_{Jup}}$ planets, 45,406 appropriate stars are in our dataset between cluster turnoff and V$=$19.5. However, this number also contains some Galactic contamination. From application of a Becanson Galactic field model \citep{Rob2003} for a WFI field in the direction of $\omega$ Cen, this contamination has been estimated at 31$\%$. By accounting for Galactic contamination in this magnitude range, we arrive at $\sim$31,000 cluster stars suitable for a search for large transiting planets. 

\subsection{Galactic Disk Contamination}
Using the Becanson model \citep{Rob2003}, we can estimate the total number of foreground Galactic disk stars in the field (non cluster members) that have photometric uncertainty $\le$2$\%$. Assuming these stars are foreground stars with solar radius or lower, they are suitable for a transit search in their own right. Cluster stars of this brightness correspond to subgiant and red giant branch members, which are unsuitable for the search. 

There are an estimated total of 6500 foreground disk stars in the field as determined from the model with $\le$2$\%$ photometry to a magnitude limit of V$=$18.2. Of these, 770 are G type main sequence stars with 160 K and 25 M main sequence stars. The remainder are of unsuitable luminosity class. By considering the small radius of M-dwarf stars, it is possible to extend a search to V$=$19.5 with sufficient photometric uncertainty to detect a planet with radius approximately equal to Jupiter. This permits a total estimate of $\sim$200 foreground M-dwarfs in the dataset. Although these numbers are insufficient for a statistical transit detection, all stars were nevertheless searched in the course of this work.

\subsection{Dataset Systematics Removal}
Before being subjected to the main transit search, the time series dataset was filtered for outlying data points and subjected to the systematics removal package of \citet{T2005}. Systematics are well known as the source of the vast majority of false transit-like features detected in large datasets, caused by various uncontrollable factors inherent in ground-based photometry (such as varying airmass and differing weather conditions). Indeed, it is clear that removal of systematics is a vital part of any transit search. The algorithm works without any previous knowledge of the particular systematics that affect the observations, and removes common trends in the data. The difference of each data point from the mean of the lightcurve is found, and the best linear fit determined. The slope is then found and the effect subtracted out, assuming the same effect is present in many lightcurves.

The algorithm does not increase the photometric precision (as seen in Fig.\space\ref{rmsplot}) except for the bright stars, whose scatter is dominated by systematics. For these the photometric precision was increased from 0.01 magnitudes to 0.003. From experimentation with the transit search algorithms, we found that the \citet{T2005} algorithm reduced the number of detected `false transits' in the data by a factor of five, while leaving transit recoverability of true artificially-induced transits unaffected.

\section{Planetary Transit Detection Algorithms}
The transit search was carried out using two separate transit detection algorithms in tandem, namely the Box fitting Least-Squares (BLS) method of \citet{K2002} and the WS method of \citet{WS2005}. The WS and BLS codes themselves are described in detail by \citet{WS2005} and \citet{K2002}, respectively, to which we direct the reader for more detailed information. In general, the matched filter approach for transit detection was first suggested by \citet{Jenk96} and has been described as the method of choice for transit searches in the general literature \citep{Ting03a,Ting03b}. Several recent experiments have adopted the method with some success (ie, \citet{G2000,Bruntt03}). 

The WS version of the matched filter method assumes a simple square-well approximation for the transit shape (a justified assumption when searching for signals near the noise level) and involves comparison of the data to a large database of suitable transit models ($\sim$10$^{6}$) with varying period, depth, duration and transit time. Each model is compared to the data via a statistical test (a cross-correlation-function in the case of \citealt{WS2005}) and a high significance output determines if a signal has been detected and at which period. By introducing detection criteria the significance of the detection `$S(P_{\rm{mod}},\tau_{\rm{shift}})$', as described in \citet{WS2005}, and the number of output data points above a pre-determined threshold (N$_{\rm{P}}$), the number of false detections can be minimized while keeping the real recoverability high.

The BLS method has also been utilized by many current transit surveys. The algorithm searches the data for the specific shape of a transit via the least-squares fitting of step-functions to the phase-wrapped data. This is repeated for a range of trial periods, fractional transit lengths, transit depths and transit epochs. The deviation of the fit to the data is calculated and the best-fitting combination of these parameters is flagged with a detection significance. This significance (SDE) is somewhat dependent on dataset properties but has been generally set at $\sim$6$\sigma$ in previous applications \citep{K2002,Moch2006}. 

Both algorithms have similar transit recoverability levels as determined via separate Monte Carlo testing on our data. The WS code was implemented with strict transit detection criteria, $S(P_{\rm{mod}},\tau_{\rm{shift}})$$\ge$11 and N$_{\rm{P}}$$\ge$10, dataset dependent values as described in \citet{WS2005} that incorporate information about the real noise in the data and the window function. The BLS code, however, produces more accurate periodicity information for candidates, leading to easier identification when phase-wrapping and searching the candidates by eye.

The degree to which the WS and BLS codes discriminate between transit-bearing artificial lightcurves and the total real dataset can be seen in Fig.\space\ref{falsedets}. The top histogram shows the BLS output SDE ($\sigma$) significance that results from running the algorithm on all dataset stars (open histogram) and on only those stars for which an artificial transit has been added via our Monte Carlo simulations (shaded histogram). The difference in the mean significance between these two populations is only marginal with a large degree of overlap between the two distributions. Hence if BLS alone was used on the dataset, most, if not all of the stars, would need to be searched with another method to identify real transit features. The apparent gap in both distributions at 5$\sigma$ is statistical, being a product of our dataset window function and the chosen bin-size.

For comparison, the bottom panel of Fig.\space\ref{falsedets} shows the WS output significance ($S(P_{\rm{mod}},\tau_{\rm{shift}})$), both for the total dataset (open histogram) and those with artificially-induced transits (shaded histogram). The difference is much more pronounced, with far less overlap, the transit-bearing stars having significantly higher output significance. By flagging candidates above a certain detection significance (marked as a solid line) far fewer stars must be further analysed to determine real transit candidates. This threshold was set at $S(P_{\rm{mod}},\tau_{\rm{shift}})$$\ge$11 for this data, as determined from tests to minimise false detection rates while keeping the recovery of transit features high. Both algorithms were therefore used in tandem, with WS used as a first-pass filter to produce a short list of candidates that were then further investigated with the BLS. 

All stars (both cluster and field) with suitable photometric precision for the transit search were analysed with WS, producing an output candidate list. These candidates were then subjected to analysis with BLS to produce more accurate period information, hence making the transits easier to see when examined by eye. No further cuts were made on these candidates. They were then visually examined for transit features by eye, both at the peak periodicity and at integer aliases thereof.

\section{Monte Carlo Simulations}
Extensive simulations with modeled transits of appropriate depth and duration were carried out with the WS and BLS algorithms to determine the transit detection efficiencies and hence expected transit recoverability in the dataset. From these recoverability results, the expected number of detected planets can be determined and information can be gained on false-detection probabilities, which should be minimised as much as possible. 

A sample of 15,307 fake transit lightcurves were produced using actual dataset lightcurves (after application of the \citet{T2005} systematics removal package) for each of three photometric uncertainty bins: stars with precision $\le$0.01 mag $\it{rms}$, 0.01$\le$mag$\le$0.02 $\it{rms}$ and 0.02$\le$mag$\le$0.04 $\it{rms}$. Over all three bins we have a total of 183,684 fake transits to test the detection algorithms. 

A box-shaped transit is superimposed on the lightcurve with a depth and duration consistent with an orbiting planet of various radii, assuming the star to be a cluster member with a previously determined radius (seen in Table.\space1). Transit models incorporating stellar limb-darkening were not produced as this effect would not affect transit visibility and/or recoverability in our data.

For the first bin (foreground stars) the stellar radius was assumed to 1.0R$_{\odot}$. For the two cluster bins, the stellar radius was determined as the weighted mean stellar radius found from the cluster theoretical isochrones. These radii are 1.23R$_{\odot}$ and 0.90R$_{\odot}$ respectively. For each depth, duration and orbital period bin, each model assumes a different epoch for the transit, spread randomly over the length of the observing run. By superimposing transits on randomly chosen dataset lightcurves, the models have the temporal resolution and photometric accuracy inherent in the real data. For the first bin (foreground stars), planets with radius 1.0R$_{\rm{Jup}}$ and 1.5R$_{\rm{Jup}}$ were simulated, whereas for the two cluster bins transits were produced that mimic the signal of planets with radius of only 1.5R$_{\rm{Jup}}$. We are insensitive to smaller cluster planets.

The highest precision bin is appropriate only for foreground disk stars (as the other stars of this brightness would be cluster giants), whereas the other two bins sample the main sequence of the cluster. The artificial transits were generated over 131 orbital periods incremented in 0.05 day steps in the range 0.52$\rightarrow$7.10d and 117 transit phases incremented by 0.2 days spanning the full MJD range of the dataset. The possibility of subday periods and a class of `Ultra Short Period' planets has been suggested by \citealt{Sahu2006}. An upper orbital period limit of seven days was used for the simulations, as this is the maximum period for which three transits may be visible over the observing run. Analysis of these artificial transits with both WS and BLS allows the sensitivity to planets of various radii to be determined, including the effects of the observational window function. 

All artificial transits were first analysed with WS, which produced a list of high significance ($S(P_{\rm{mod}},\tau_{\rm{shift}})$$\ge$11) preliminary candidates. These candidates were then analysed with BLS for more accurate period determination. Those recovered transits all had BLS-determined periods within $\pm$0.1d of the real injected period. 

Examples of artificial transits that were detected at the low end of the detection criteria (SDE $\sim$6.0$\sigma$ significance with BLS, $S(P_{\rm{mod}},\tau_{\rm{shift}})$$\ge$11 with WS) can be seen in Fig.\ref{fakesplot}, indicating the visibility of transits in the dataset when phase-wrapped to the peak detected period. The panels are arranged in order of increasing photometric uncertainty, with typical orbital periods of VHJ (1.5d, left panels) and HJ (3.3d, right panels). 

The top two panels of Fig.\space\ref{fakesplot} show a transit (at $\Phi=$0 and 1.0) for a 1.0R$_{\rm{Jup}}$ planet transiting a star of 1$\%$ photometry, typical for the transit visibility for the brighter foreground stars. Moving downward, the two pairs of panels show a transit of a 1.5R$_{\rm{Jup}}$ planet and a star of 2$\%$ photometry, and a transit of 1.5R$_{\rm{Jup}}$ against a 4$\%$ star, which defines the bottom of our search regime. The bottom two sets of panels are appropriate for stars on the cluster main sequence.

Fig.\space\ref{mchist} shows the resulting transit recoverability (as a fraction of the total number of modeled transits per period bin) for those transits for stars with $\le$0.01 mag $\it{rms}$ (foreground stars, top left) and 0.01$\le$mag$\le$0.02 $\it{rms}$ (top right). These transits are those that passed both the $S(P_{\rm{mod}},\tau_{\rm{shift}})$$\ge$11 and N$_{\rm{P}}\ge$10 detection criteria of the WS, and had periods within 0.1d of the true injected transit period (although this in itself is not a criteria that a candidate must pass). These criteria were identical to those applied to the real data. For the brightest stars it can be seen that the recoverability is good to P$_{\rm{orb}}$$=$2d for those planets with radii $\ge$1.0R$_{\rm{Jup}}$. A large drop in recoverability can be seen for periods approaching one day for all planetary radii, caused by terrestrial effects.

As the photometric precision of the data becomes worse, only transits corresponding to larger planets can be detected. For stars with 0.01$\rightarrow$0.04 mag $\it{rms}$, the recoverability of planets with R~$<$~1.5R$_{\rm{Jup}}$ becomes increasingly truncated. By considering this, as well as the expected transit depths as a function of V magnitude in the cluster (see Fig.\space\ref{rmsplot}), the search can only provide meaningful statistics on planets with a radius of 1.5R$_{\rm{Jup}}$. The chosen upper limit is arbitrary, but is consistent with the upper limit of currently known transiting planets, particularly the discoveries of HAT-P1-b and WASP-1b \citep{Bakos2006,CC2006} that are approaching this radius.

The bottom left panel of Fig.\space\ref{mchist} displays the 1.5R$_{\rm{Jup}}$ transit recoverability for stars with photometric precision between 2$\rightarrow$4$\%$ (0.02$\rightarrow$0.04 mag). These stars correspond to those with V$=$18.5$\rightarrow$19.5 on the cluster main sequence. This defines the lower magnitude limit of the search, as the transit depth becomes smaller than the photometric scatter below this point. The recoverability for these stars is good for P$_{\rm{orb}}\le$1.5d. The bottom right panel of Fig.\space\ref{mchist} displays the weighted mean transit recoverability for all simulations, taking into account the relative numbers of stars that lie in each photometric bin. These numbers were used as part of the determination of the expected number of planets detectable in the cluster.

Fig.\space\ref{magnumplot} shows the transit recoverability for 1.5R$_{\rm{Jup}}$ planets as a function of V magnitude on the cluster main sequence, assuming that they have short periods. The left panel depicts the recoverability of VHJ planets (with an assumed orbital period of 1.5d); the right panel shows the same for HJ planets (assumed period of 3.3d). The rapidly increasing stellar radii past the MSTO defines the bright limit to the recoverability, whereas the increasing photometric uncertainty (Fig.\space\ref{rmsplot}) drives the faint limit. It can be seen that for VHJ, the recoverability is good for stars 18.0$\le$V$\le$19.0. For HJ planets with R~$<$~1.5R$_{\rm{Jup}}$, the recoverability is low, as this is the lower limit to the planetary radius for which we can search at such faint magnitudes (Fig.\space\ref{rmsplot}).

\section{Sensitivity to Short-Period Planets}
With the results of the Monte Carlo simulations, we now determine our 
sensitivity to planets in $\omega$ Cen considering a number of factors, 
including the total number of stars monitored with the appropriate 
photometric precision, the transit probability and transit recoverability (as determined via the Monte Carlo simulations) as a function of planetary period, and 
the assumed distribution of planets across orbital period. 
This sensitivity allows us to assess the significance of the results 
from our planet search, which we conduct for short-period planets with 
orbital periods less than seven days, and thus encompassing both 
Very Hot Jupiter (VHJ) and Hot Jupiter (HJ) exoplanets. The essential 
question to ask is, ``For how many stars would we have detected 
a transit signal if they were hosts to short-period planets?''  
In this section we answer that question, and tabulate the result under 
different assumptions in Table~2.

Since both the probability that an exoplanet will transit from the viewer's 
perspective and the 
probability that it will be detected with our transit algorithm are 
dependent on the planetary orbital period, an assumption must be made 
about how VHJ and HJ planets are distributed across period.  
We consider four different scenarios, VHJ planets distributed 
equally over the range 1.0$\le$P$<$3.0d (VHJ); HJ planets distributed 
equally over either 3.0$\le$P$<$5.0d (HJ5) or over 3.0$\le$P$<$7.0d (HJ7); and 
short-period planets distributed equally over 1.0$\le$P$<$5.0d periods (VHJ+HJ).
For each hypothesis, the 31,000 stars in $\omega$~Cen 
for which we have suitable precision to detect transiting planets are 
uniformly apportioned to the relevant period bins, and given in 
column 5 of Table~2 as N$_{*,\rm{mon,i}}$.  
 
Not all of these planets will transit their host stars, so that the transit probability must be factored in. This probability, P$_{\rm{trans,i}}$, depends on the planetary orbital period and radius of the host star, via:
\vspace{-0.2cm}
\begin{center}
\begin{equation}
P_{\rm{trans,i}} \sim \big(\frac{R_{\ast}}{\alpha}\big)
\end{equation}
\end{center}
\noindent where $R_{\ast}$ is the stellar radius and $\alpha$ is the planetary semi-major axis. 

The radius of the host star ($R_{\ast}$) varies with V magnitude in the cluster, hence a weighted mean radius for the transit search stars was used in the calculation. By considering the stellar radius from 17.5$\le$V$\le$19.5 (Table\space1) in bins of half a magnitude, as well as the relative fraction of the total sample present in each bin, the weighted mean stellar radius was calculated to be 0.96$\rm{R_{\odot}}$. The transit probability was then found for each planetary orbital period P$_{i}$, and is displayed in column 3 of Table\space2.
 
The final factor to consider is the probability that our algorithm will detect 
or recover the transit. Determined from the Monte Carlo simulations, this recoverability, R$_{\rm{trans,i}}$, incorporates the window function and the true noise characteristics of the data.  It is calculated for each planetary radius and 
the middle of each period bin from the weighted mean transit recoverability (bottom right panel of Fig.\space\ref{mchist}). These values are 
given in the fourth column of Table\space2 assuming a planetary radius of 1.5R$_{\rm{J}}$.

The total number of equivalent stars, N$_{\rm{*,probe}}$, probed at full sensitivity for planets distributed according to one of our four hypothesis is then given by 
the simple sum of the numbers recorded in column 6 of Table~2, ie, N$_{\rm{*,probe}}=$$\sum_{i}$N$_{\rm{*,mon,i}} \times $P$_{\rm{trans,i}} \times $R$_{\rm{trans,i}}$. 
Fig.\space\ref{expnums} shows this proxy for sensitivity, N$_{\rm{probe,i}}$, 
as a function of orbital period for each of the four hypotheses and assuming a planetary radius of 1.5R$_{\rm{Jup}}$ in each case. Smaller planets will escape detection in our $\omega$ Cen experiment.

\section{Search Results}
Using the same search methods as in the Monte Carlo simulations, all 45,406 (31,000 cluster) stars with suitable photometric accuracy ($\le$0.04 mag) in our $\omega$ Cen dataset were first analysed with the WS algorithm, and a candidate list produced from those lightcurves that displayed a $S(P_{\rm{mod}},\tau_{\rm{shift}})$$\ge$11 output significance and a N$_{\rm{P}}$$\ge$10 criteria from the WS algorithm (see Fig.\space\ref{falsedets}). 

After filtering for detections with common inherent features (ie, those detections that occur at the same times and with the same periods, within 0.001d), 138 candidates remained. These candidates were analysed with BLS and their peak periodicities determined. They were then phase-wrapped to this period (and integer aliases) and visually examined for transit features. 

The nature of the periodicity was immediately seen from the phase wraps. The vast majority of the candidates had transit-like features that could be attributed to bad columns in the CCDs, to non-perfect PSF fitting to the star due to blended companions, or association with stars close to the dataset saturation limit. Some candidates were composed of single outlying data points and others of random groupings of data points close to integer days which had no apparent pattern. Candidates with such features were classified as false positives.

After this culling, none of the candidates (either in the foreground or the cluster) displayed transits of suitable depth and duration that could be attributed to an orbiting giant planet in our radius regime. The list did, however, include 8 eclipsing binaries and 2 $\delta$ Scuti stars that were previously identified and published in \citet{W2006}. 

The lack of planetary detections cannot be attributed to 
our algorithms, as evidenced by the Monte Carlo simulations and a 
control field in the Lupus Galactic Plane that we observed for 53 nights during 2005 and 2006 with the same instrument and detector. The Lupus data were analyzed with the same methods as used for $\omega$ Cen. Several transit candidates were found in Lupus, including one candidate which has excellent prospects for being a transiting HJ planet \citep{W2007a}. No similar candidates were seen in $\omega$ Cen.

The net result of our search for transiting planets in 
$\omega$~Cen is thus: no planetary candidates, 10 variable stars (previously 
detected with our variable star search algorithm), and 128 false positives.  
In the next subsection, we examine the implications of our null detection of large, short-period planets in the cluster.  

\subsection{Upper Limits to Planetary Frequency in $\omega$ Cen}
An upper limit to the occurrence frequency $F$ of large-radius VHJ and HJ planets in $\omega$ Cen can be determined using the results of our survey. Using Poisson statistics to analyze the significance of 
our null result under each of our four distribution hypotheses, we place an upper limit 
on the fraction of stars in $\omega$ Cen that are orbited by short-period 
planets.  For example, VHJ planets with 1.0~$\le$~P~$<$~3.0d must 
have a frequency of occurence $F_{\rm VHJ} < 1/1040$ in order for 
none to be detected 95$\%$ of the time 
in our sample of N$_{\rm *,probe} = $3100 ``equivalent 
stars'' probed at full sensitivity.

Similarly, at 95$\%$ confidence, the frequency of HJ planets in $\omega$ Cen 
is $F_{\rm HJ5} < 1/150$  or $F_{\rm HJ7} < 1/93$, depending on whether they 
have periods P that are equally distributed over $3 - 5$~days or $3 - 7$~days, 
respectively. If there is no strong difference in occurrence frequency between VHJ and HJ, 
so that we may assume that the combined population is equally distributed 
over 1d~$< P < $5d, the upper limit on the frequency of such short-period planets is 
$F_{\rm VHJ+HJ} < 1/600$ (95$\%$CL). These upper limits are tabulated in Table~2.

\subsection{Metallicity considerations}
In the Solar Neighborhood, evidence suggests that planetary frequencies are 
influenced by the metallicity of the host star \citep{G1997,L2000,S2001,FV2005}, 
as also appears to be the case in the globular cluster 47 Tucanae \citep{W2005}. Based on statistics gathered in radial velocity searches, \citet{M2005} conclude that the probability, P$_{\rm{planet}}$, that a given star will host a planet is related to metallicity via P$_{\rm{planet}}=$0.03$\times$10$^{2.0 {\rm [Fe/H]}}$, where [Fe/H] is the stellar metallicity and the constant of proportionality is assumed to be the same for planets of all orbital periods. At very low metallicities, the relation is poorly constrained observationally, and thus highly uncertain. 

Of the 31,000 monitored cluster stars in $\omega$ Cen, 80$\%$ are assumed to be members of the main metal-poor population, with corresponding [Fe/H]$=$$-$1.7, taken from the results of \citet{N2004}. Similarly, 15$\%$ and 5$\%$ are assumed to have [Fe/H]$=$$-$1.2 and $-$0.6 respectively. The weighted mean metallicity for the sampled stars is hence [Fe/H]$=-$1.57.

If the above relation between P$_{\rm{planet}}$ and stellar metallicity also holds true for globular clusters, then the probability that planets will be found around stars in $\omega$ Cen is reduced, by an amount that depends on the extrapolation of 
the relation into uncertain territory.  A conservative estimate, that we will 
assume in the next section, is that 
$\omega$ Cen planetary occurrence frequency is suppressed by at least the $\sim$60$\%$ reduction in $F$ estimated by \citet{M2005} for exoplanets 
around stars of [Fe/H]$=$$-0.5$ compared to those with nearly solar metallicities. In the next section, we compare our results 
with rates and frequency limits of short-period planets in non-cluster 
Galactic environments.

\section{Discussion and Implications}
We have placed upper limits on the frequency with which $\omega$ Cen stars 
host short-period (VHJ and HJ) planets.  How do these limits compare to the 
frequency of VHJ and HJ planets detected in other Galactic fields? 

\citet{Gouldetal2006} analyzed the OGLE III transit surveys in 
Galactic disk and bulge fields to derive occurrence rates for 
VHJ and HJ planets with radii satisfying 
1.0~$<$ R$_{\rm{Jup}} <$~1.25.  These rates, according to the authors, 
are statistically indistinguishable from those derived from radial velocity 
results on bright nearby stars. No larger, short-period planets 
were detected by the OGLE~III surveys despite increased sensitivity, 
leading \citet{Gouldetal2006} to compute an upper limit $F$ for these larger worlds.

For Very Hot Jupiter (VHJ) planets uniformly distributed over 1d~$<$~P~$<$~3d, 
the \citet{Gouldetal2006} result is $F = $1/710(1$^{+1.10}_{-0.54}$) at a 90$\%$ confidence level. Their 95\% confidence rate is thus somewhat larger than our 
upper limit of $F_{\rm VHJ} < 1/1040$ for 
somewhat larger planets in $\omega$ Cen, possibly indicating suppression 
in the cluster.  However, within the 90$\%$ confidence 
envelope, the two are consistent regardless of whether a correction 
for metallicity is applied or not.

For Hot Jupiter (HJ) planets uniformly distributed over 3d~$<$~P~$<$~5d, 
\citet{Gouldetal2006} find $F =$~1/320(1$^{+1.37}_{-0.59}$) at a 90$\%$ confidence 
for smaller transiting planets in OGLE~III fields.  For larger planets 
in $\omega$~Cen, our upper limit is $F_{\rm HJ5} < 1/150$ using the same 
assumption about period distribution.  Since the upper limit on HJ planets in 
this cluster environment is larger than the occurrence rate 
inferred for OGLE fields, even without a metallicity correction, 
no conclusion can be drawn. 

Perhaps the most direct comparison that can be made of the two studies 
is the upper limits on the frequency of $~R=$~1.5 R$_J$ planets 
assuming uniform distribution over the orbital 
period range 1d~$<$~P~$<$5d.  Our result for $\omega$~Cen is 
$F_{\rm VHJ+HJ} < 1/600$.  The corresponding value for OGLE~III 
fields can be inferred from Table~6 of \citet{Gouldetal2006} to 
be a very comparable upper limit of $F < 1/640$.  

\citet{Freg2006} (and earlier, \citealt{Sig1992}) carried out a study of dynamical interactions of planet-bearing stars with other stars in dense stellar environments via N-body simulations. They conclude that planet survivability is greater in dense systems than had been previously predicted, and presented a characteristic timescale for planet survivability for various star-planet-star mass ratios, systemic stellar densities and velocity dispersions (see Fig.\space6 and Eq\space22 in \citealt{Freg2006}).

$\omega$ Cen has a measured core stellar density of only 1/10$^{\rm{th}}$ the core density of 47 Tuc, and a measured core velocity dispersion of 20km/s \citep{MM86,VDV2006}, which falls to 14km/s at the cluster half-mass radius. Using these parameters, the expected lifetime for a planetary system to remain bound in the cluster is $>$1$\times$10$^{10}$ years. 

\citet{Sig1992} has shown that the survivability is long ($\sim$10$^{8}$yrs) for a planet of $\sim$1AU semi-major axis in an environment typical of the core of 47 Tuc, and even longer for the lower density environment of $\omega$ Cen. For the short period HJ planets, the probability of disruption due to stellar interaction is very low. The transiting planet search presented here is most sensitive to the outer halo of the cluster, where stellar densities are far lower than in the core. Hence 1$\times$10$^{10}$ years can be taken as a firm lower limit to the timescales of planet survivability in the cluster. 

The conclusion is that if the planets formed in the first place, then they are expected to survive in $\omega$ Cen through its current and remain detectable. The low upper 
limits on the frequency of short-period planets in the cluster are consistent 
with its low metallicity inhibiting planet formation from the outset.

Our null result does not rule out the existence of large-radius planets in longer period orbits in the cluster. The confirmation of a planetary mass object in the globular cluster M4 \citep{Sig2003} provides the first direct evidence of planetary formation in a very metal poor environment. The paucity of transiting 
short-period large planets in globular clusters is due to a process other 
than stellar dynamics, perhaps a dependency of planetary migration on stellar metallicity. 

\section{Summary and Conclusions}
We have presented the results of a wide-field, deep photometric search for transiting short-perod planets in the globular cluster $\omega$ Centauri, a region previously un-sampled for planetary transits. The cluster was observed with a 52$'$$\times$52$'$ field of view for 25 contiguous nights with the ANU 40-inch telescope at Siding Spring Observatory. From application of difference imaging analysis, a total of 109,726 time series were produced across the field, each being composed of 787 independent data points.

A total database of 45,406 stars have photometric accuracy suitable for the search ($\le$0.045 mag scatter down to V$=$19.5), including 31,000 cluster stars extending 2.5 magnitudes down the main sequence. All of these were subjected to a rigorous 
(and vigorous) search for transit-like events; none were detected after variable stars and clear false-positives 
were removed.  Simulations have shown that if large Hot Jupiters (HJs) formed in the cluster then dynamically speaking they would survive to be detectable in our data.

Extensive Monte Carlo simulations via injection of transit signals into actual 
light curves were used to determine the sensitivity of 
the survey to R$\le$1.5R${\rm_{Jup}}$ planets over a range of 
orbital periods.  Coupled with our null result, we are thus able to place strict, 
statistically significant upper limits on the occurrence frequency $F$ of 
large (R~$=$~1.5R), short-period planets in $\omega$ Centauri.  

We determine a limit of $F_{\rm VHJ} < 1/1040$ at 95$\%$ confidence 
for Very Hot Jupiter (VHJ) planets with periods distributed 
uniformly over 1d~$<$~P~$<$~3d. This upper limit for the cluster is less 
than that determined by \citet{Gouldetal2006} for smaller (1.3R${\rm_{Jup}} < $~R~$< 1.5$R${\rm_{Jup}}$) planets with the same period distribution in the Galactic fields 
surveyed by OGLE~III.  The two results are consistent at the 90$\%$ confidence 
level, and more understandable if the low metallicity of $\omega$ Cen 
suppresses planet formation or planetary migration.

Under the assumption that there is no difference in occurrence 
frequency for VHJ and HJ across the orbital period range 
1d~$<$~P~$<$5d, we derive an upper limit of $F_{\rm VHJ+HJ} < 1/600$ in $\omega$ Cen.
The corresponding result in the Galactic OGLE~III fields for 
comparably-sized planets is an upper limit of 
$F < 1/640$.  Both results are quoted at 95$\%$ confidence.  Our results are 
less dependent on model assumptions about the distance to the target 
population since the vast majority of stars in our fields are members of, and 
thus at the distance of, $\omega$~Cen. 

It is noteworthy that despite the fact that the OGLE~III campaigns monitored considerably more stars in better median seeing with more frames per field, our $\omega$ Cen study produces a comparable upper limit on the frequency of large, short-period planets. While part of the reason may lie in the longer exposure times, denser sampling, and the use of different cleaning and detection algorithms in our survey, a large part of the difference is due to the small fraction of the more distant and obscured Galactic bulge stars (which constituted about a third of the total OGLE~III sample) that can be meaningfully probed for transiting planets.  

This null result for VHJ and HJ planets in $\omega$ Cen, coupled with the null result of 47 Tucanae \citep{W2005} strengthens the evidence for the dominance of system metallicity over stellar interactions in determining short period planetary frequencies in globular clusters. At longer orbital periods stellar encounters may play a role in determining planetary frequencies. This is a result aligned with current work on the metallicity trend of planet-bearing host stars in the Solar Neighborhood and N-body simulations of planets in dense environments. Such a metallicity dependence is one of the main predictions of the core accretion model of planet formation. 

\acknowledgements
The authors would like to thank Omer Tamuz for discussions on the removal of dataset systematics, Laura Stanford for help with the production of the theoretical cluster isochrones, and the referee for a very thorough and helpful review process.

\clearpage

\begin{table}
\begin{center}
\caption{}
\footnotesize
\begin{tabular}{llllllll} \hline
\noalign{\medskip}
$\rm{V}$ & $\rm{Mass(M_{\odot})}$ & $\rm{T(K)}$ & $\rm{L(L_{\odot})}$ & $\rm{log(g)}$ & $\rm{M_{V}}$ & $\rm{R(R_{\odot})}$ & $N_{\ast}$\\
\noalign{\medskip}
\hline
\noalign{\medskip}
17.0 & 0.7636 & 5500 & 4.08 & 3.6212 & +3.4 & 2.24 & 1300\\
17.25 & 0.7613 & 5700 & 3.21 & 3.7848 & +3.65 & 1.85 & 1700\\
17.5 & 0.7579 & 6000 & 2.64 & 3.9375 & +3.9 & 1.55 & 2200\\
17.75 & 0.7487 & 6100 & 1.89 & 4.1312 & +4.15 & 1.23 & 2800\\
18.0 & 0.7426 & 6100 & 1.62 & 4.2003 & +4.4 & 1.13 & 3200\\
18.25 & 0.7272 & 6100 & 1.18 & 4.3167 & +4.65 & 0.98 & 3060\\
18.5 & 0.7186 & 6000 & 1.01 & 4.3653 & +4.9 & 0.92 & 2900\\
18.75 & 0.6993 & 5900 & 0.76 & 4.4516 & +5.15 & 0.82 & 2700\\
19.0 & 0.6897 & 5900 & 0.65 & 4.4917 & +5.4 & 0.78 & 2500\\
19.25 & 0.6687 & 5800 & 0.49 & 4.5629 & +5.65 & 0.71 & 2200\\
19.5 & 0.6575 & 5700 & 0.43 & 4.5945 & +5.9 & 0.68 & 2000\\
20.0 & 0.6092 & 5400 & 0.26 & 4.6978 & +6.4 & 0.58 & 1800\\
\noalign{\medskip}
\hline
\noalign{\medskip}
\hline\hline
\end{tabular}  
\end{center}
\tablecomments{The determined stellar parameters for $\omega$ Cen turnoff and main sequence stars from V$=$17.0 to V$=$20.0 as produced from \citet{Y2003} theoretical isochrones. The tabulated values are the apparent V magnitude (using the best-fit distance modulus of 13.6 and a metallicity of $-$1.7dex), the stellar mass in solar units, the stellar temperature and luminosity in solar units, the logarithm of the surface gravity, the corresponding absolute magnitude and the determined stellar radius along with the estimated number of cluster stars present in each magnitude bin. These radius considerations were used in determining transit recoverability and visibility.}
\end{table}

\clearpage
\thispagestyle{empty}
\setlength{\voffset}{-20mm}
\begin{table}
\begin{center}
\caption{}
\footnotesize
\begin{tabular}{lllllr} \hline
\noalign{\medskip}
$\rm{\it{P}_{i}}$ (days) & $\rm{Rel~P_{occur,i}}$ & $\rm{P_{trans,i}}$ & $\rm{R_{trans,i}}$ & $\rm{N_{*mon,i}}$ & $\rm{N_{*probe,i}}$ \\
\noalign{\medskip}
\hline
\noalign{\medskip}
0   & 0 & - & - & - & -  \\
0.5 & 0 & - & - & - & -  \\
1.0 & 0.25  & 0.20 & 0.84 & 7750 & 1302\\
1.5 & 0.25  & 0.18 & 0.70 & 7750 &  977\\
2.0 & 0.25  & 0.15 & 0.43 & 7750 &  500\\
2.5 & 0.25  & 0.14 & 0.31 & 7750 &  336\\
\noalign{\medskip}
\hline
 & & & & & $\rm{N_{*,probe}}=$3100\\ 
 & & & & & $F_{\rm{VHJ}} < $1/1040\\ 
\noalign{\medskip}
3.0	& 0.25	& 0.12	& 0.19	& 7750 & 177\\
3.5	& 0.25	& 0.11	& 0.16	& 7750 & 136\\
4.0	& 0.25	& 0.10	& 0.13	& 7750 & 101\\
4.5	& 0.25	& 0.09	& 0.07	& 7750 &  49\\
\noalign{\medskip}
\hline
 & & & & & $\rm{N_{*,probe}}=$460\\ 
 & & & & & $F_{\rm{HJ5}} < $1/150\\ 
\noalign{\medskip}
3.0 & 0.125 & 0.12 & 0.19 & 3875 &   88.4\\
3.5 & 0.125 & 0.11 & 0.16 & 3875 &   68.2\\
4.0 & 0.125 & 0.10 & 0.13 & 3875 &   50.4\\
4.5 & 0.125 & 0.09 & 0.07 & 3875 &   24.4\\
5.0 & 0.125 & 0.09 & 0.05 & 3875 &   17.4\\
5.5 & 0.125 & 0.09 & 0.05 & 3875 &   17.4\\
6.0 & 0.125 & 0.08 & 0.04 & 3875 &   12.4\\
6.5 & 0.125 & 0.07 & 0.02 & 3875 &    5.4\\
\noalign{\medskip}
\hline
 & & & & & $\rm{N_{*,probe}}=$280\\ 
 & & & & & $F_{\rm{HJ7}} < $1/93\\ 
\noalign{\medskip}
0   & 0 & - & - & - & -  \\
0.5 & 0 & - & - & - & -  \\
1.0	& 0.125	& 0.20	& 0.84	& 3875	& 651\\
1.5	& 0.125	& 0.18	& 0.70	& 3875	& 489\\
2.0	& 0.125	& 0.15	& 0.43	& 3875	& 250\\
2.5	& 0.125	& 0.14	& 0.31	& 3875	& 168\\
3.0	& 0.125	& 0.12	& 0.19	& 3875	&  88.4\\
3.5	& 0.125	& 0.11	& 0.16	& 3875	&  68.2\\
4.0	& 0.125	& 0.10	& 0.13	& 3875	&  50.4\\
4.5	& 0.125	& 0.09	& 0.07	& 3875	&  24.4\\
\noalign{\medskip}
\hline
 & & & & & $\rm{N_{*,probe}}=$1800\\ 
 & & & & & $F_{\rm{VHJ+HJ}} < $1/600\\ 
\noalign{\medskip}
\hline\hline
\noalign{\medskip}
\end{tabular}  
\end{center}
\tablecomments{Numerical parameters used to calculate the total number of stars 
probed for the four assumptions described in the text, for the distribution of 
VHJ planets as a function of period.  The tabulated parameters as described in section 7 are the orbital period for bin $\rm{i}$, $\rm{\it{P}_{i}}$, the transit probability, $\rm{P_{trans,i}}$ the detection algorithm transit recoverability, $\rm{R_{trans,i}}$, for that bin, the number of stars monitored in the bin, $\rm{N_{*mon,i}}$, 
and the resulting number of stars probed $\rm{N_{*probe,i}}$ in $\omega$ Cen. 
Since the total number (31,000) of $\omega$ Cen stars monitored with 
sufficient photometric precision is estimated to two significant figures, $\rm{N_{*probe,i}}$ is similarly estimated. For each hypothesis, an upper limit (95$\%$ CL) 
$F$ to the occurrence frequency of short-period 
planets is also given.}
\end{table}
\clearpage
\setlength{\voffset}{0mm}

\figcaption[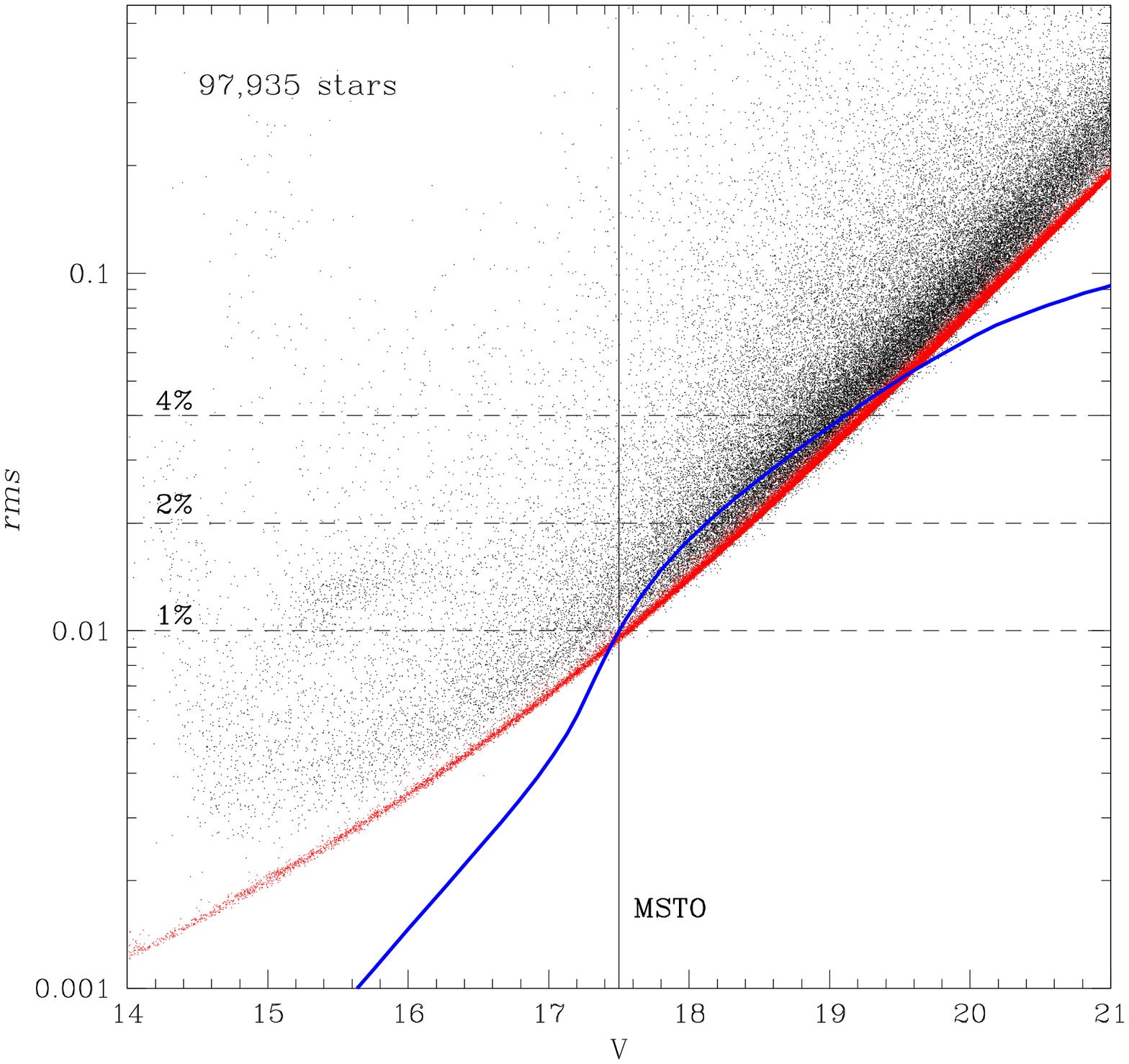]{The measured photometric $\it{rms}$ uncertainty after application of systematics removal for 97,935 stars in the $\omega$ Cen field that were cross-identified with the CMD dataset, measured as the standard deviation as a function of V magnitude. Also plotted as the red points is the total noise contribution (star+sky+residual noise) calculated for each star independently from their respective star and sky fluxes. Also plotted is the location of the cluster MSTO (at V$=$17.5). Various percentage uncertainties are marked for clarity. It can be seen that the photometric uncertainty is 0.01 mag at the cluster turnoff, degrading to 0.04 mag at V$=$19.0. Also overplotted are the expected transit depths of a planet with R$=$1.5R$_{\rm{Jup}}$ for comparison with the photometry. Stars as faint as V$=$19.5 can be searched for R$=$1.5R$_{\rm{Jup}}$ planets with a transit depth equal to the photometric uncertainty. We are insensitive to planets with smaller radii.\label{rmsplot}}

\figcaption[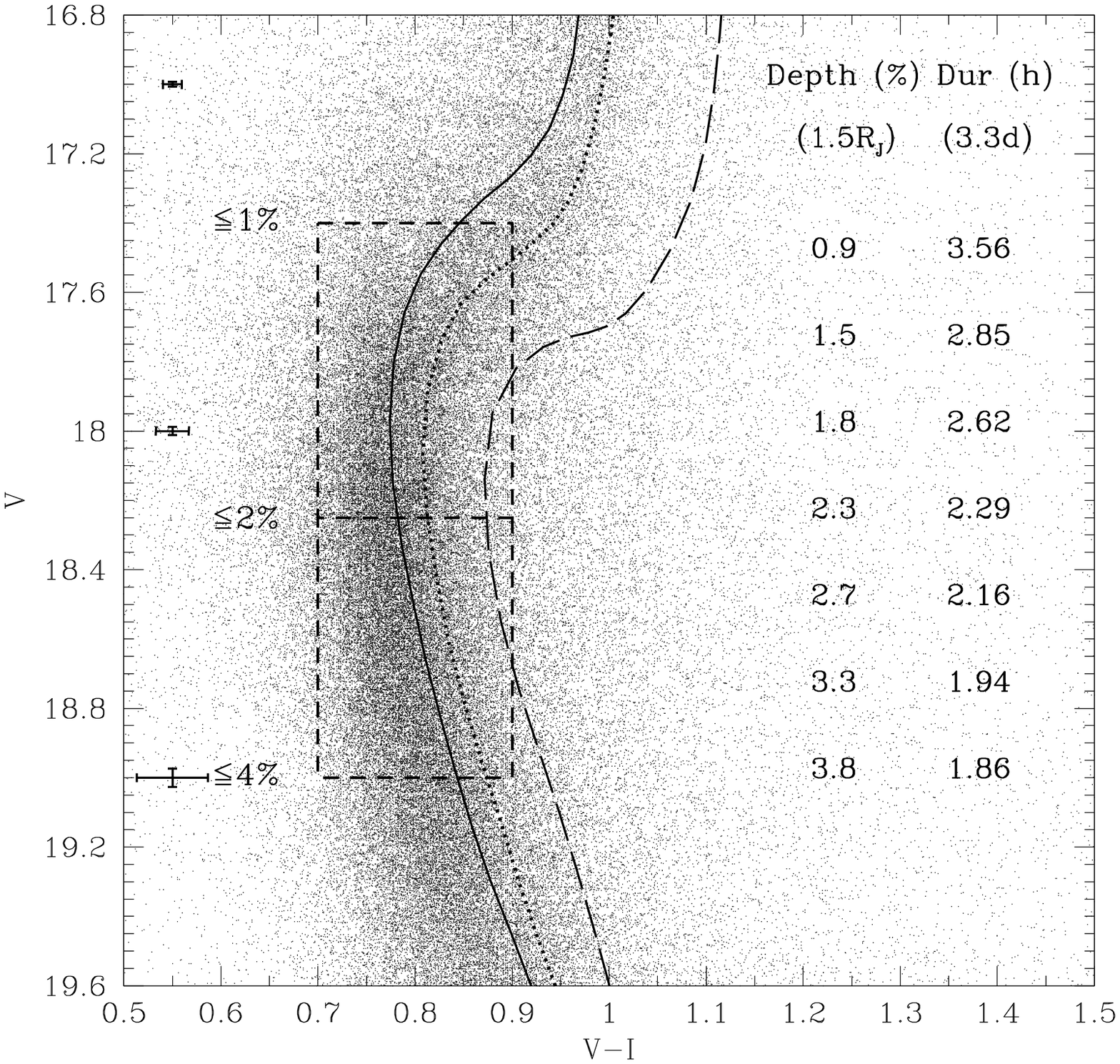]{The color magnitude diagram for the upper main sequence of $\omega$ Cen. The boxes contain the stars searched for transits based on the depth of the expected signal compared to the photometric precision. Overplotted are three \citet{Y2003} theoretical isochrones to describe the cluster population with [Fe/H]$=$$-$1.7 (solid line),$-$1.2 (dotted line) and $-$0.8 (dashed line), values as taken from \citet{N2004}. Assuming the best-fitting distance modulus of 13.6, the radius and mass of the main sequence stars were found as a function of V, and the corresponding depths (in magnitude units) and durations (in hours) of an orbiting 1.5 R$_{\rm{Jup}}$ planet were calculated and overplotted. The Monte Carlo simulations were carried out with these parameters. The duration was calculated assuming an orbital period of 3.3d, typical for HJs in the solar neighborhood.\label{transcmd}}

\figcaption[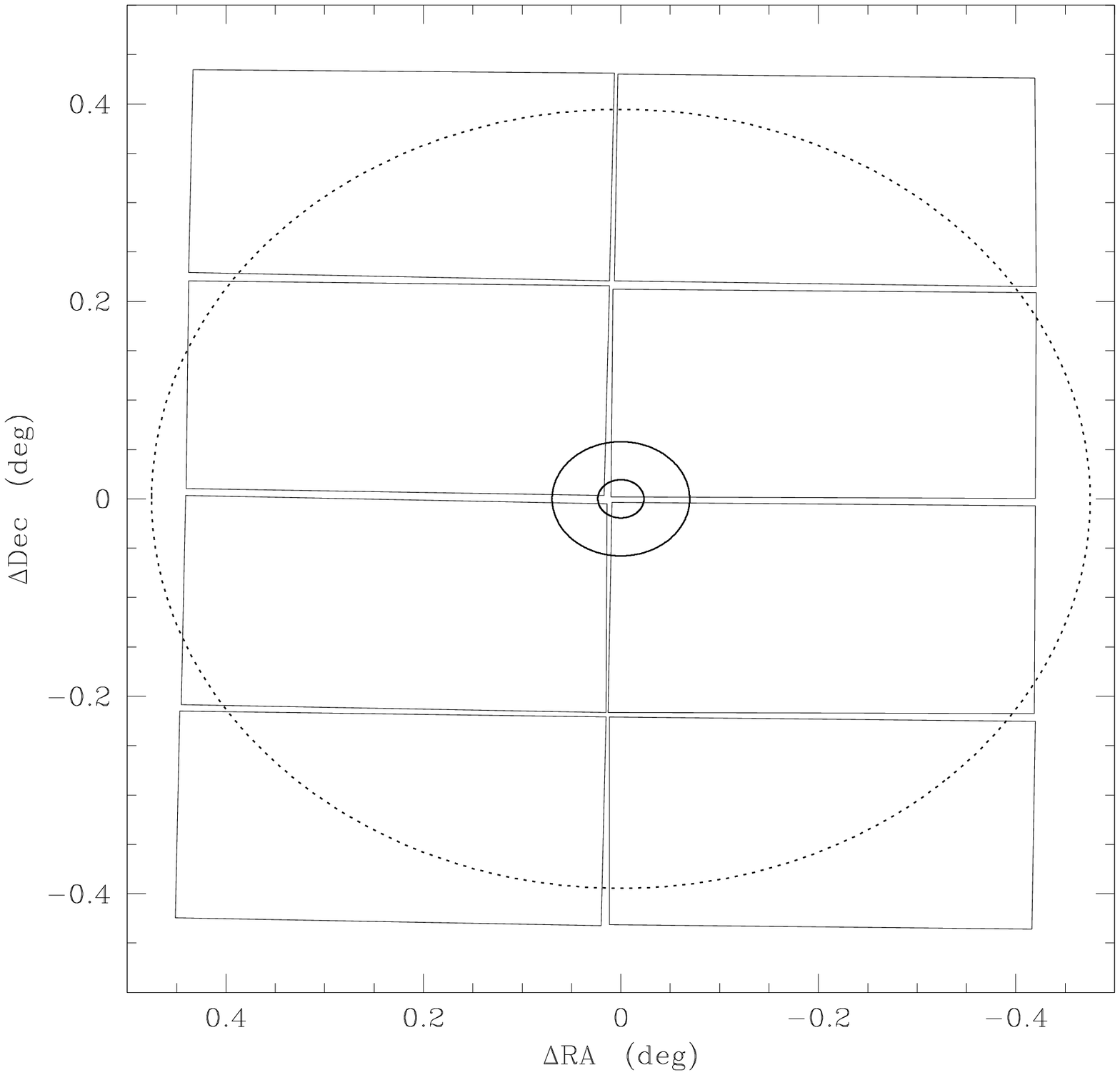]{The extent of our WFI field of view, shown as $\Delta$RA and $\Delta$Dec (in degrees) from the core of the cluster. The locations of the eight WFI CCDs are overplotted with the locations of the cluster core radius (inner ellipse), the cluster half-mass radius (middle ellipse) and the position of 50$\%$ of the cluster tidal radius (outer ellipse). Cluster parameters are taken from \citet{Harris96}.\label{chips}}

\figcaption[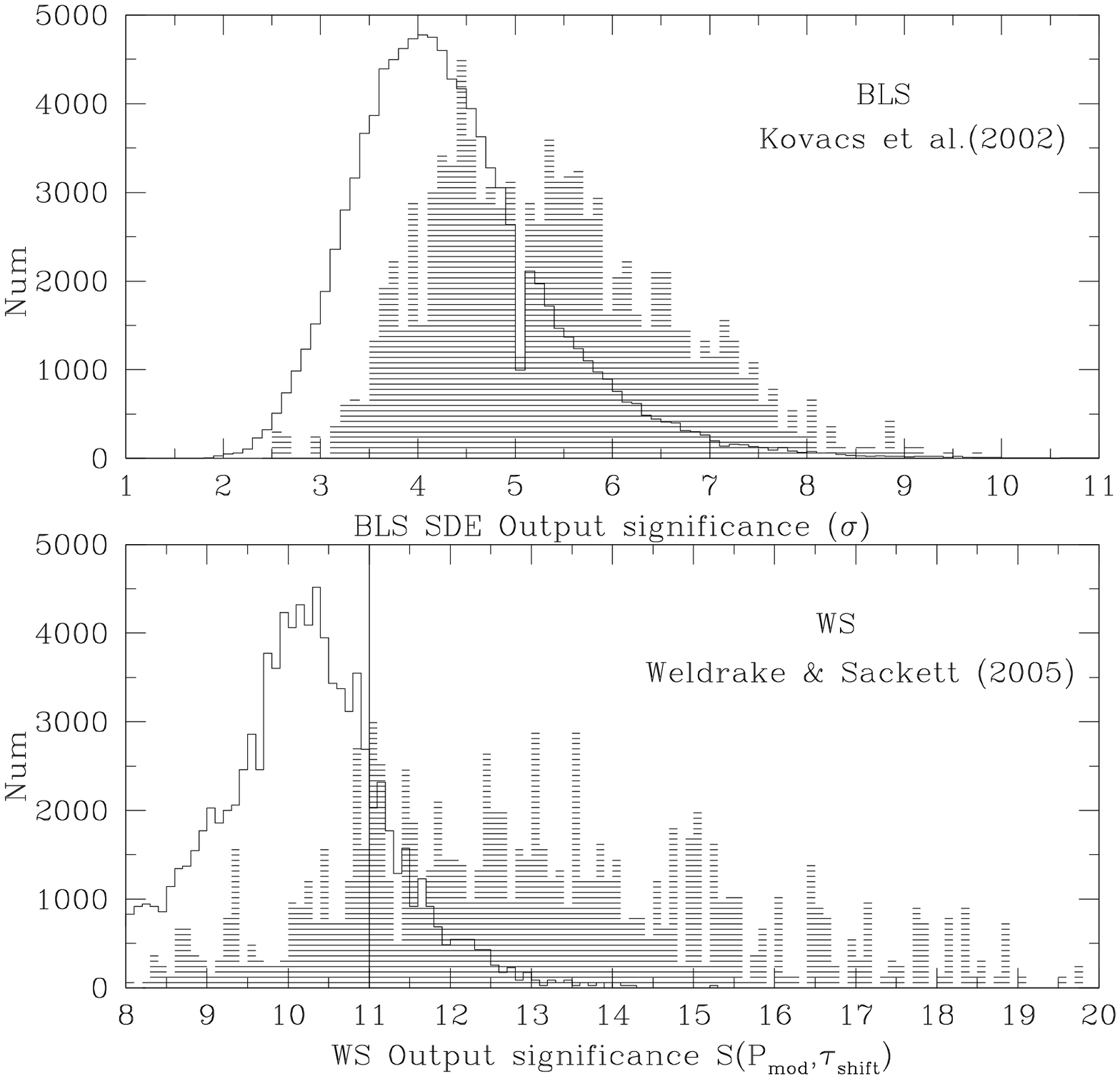]{A comparison of the degree to which the BLS (top) and WS (bottom) detection codes discriminate between lightcurves with artificially induced transits (dark shading) and no added transits (open histogram) superimposed on our actual observational light curves. The histograms of significances for the BLS code are similar whether transits are present or not, hence most, if not all, of the stars must be examined by eye at the peak detected period to identify the transits. The WS code, however, has more disparate histograms for the two populations, leading to far fewer false detections in the output candidate lists. The BLS code produces more accurate period determinations. Hence it was decided to use both algorithms in tandem: the WS to first filter the lightcurves for high significance candidates (with a threshold marked by the solid line and an N$_{\rm{P}}$ criterion of $\ge$10 points) and the BLS peak detection for accurate period determination.\label{falsedets}}

\figcaption[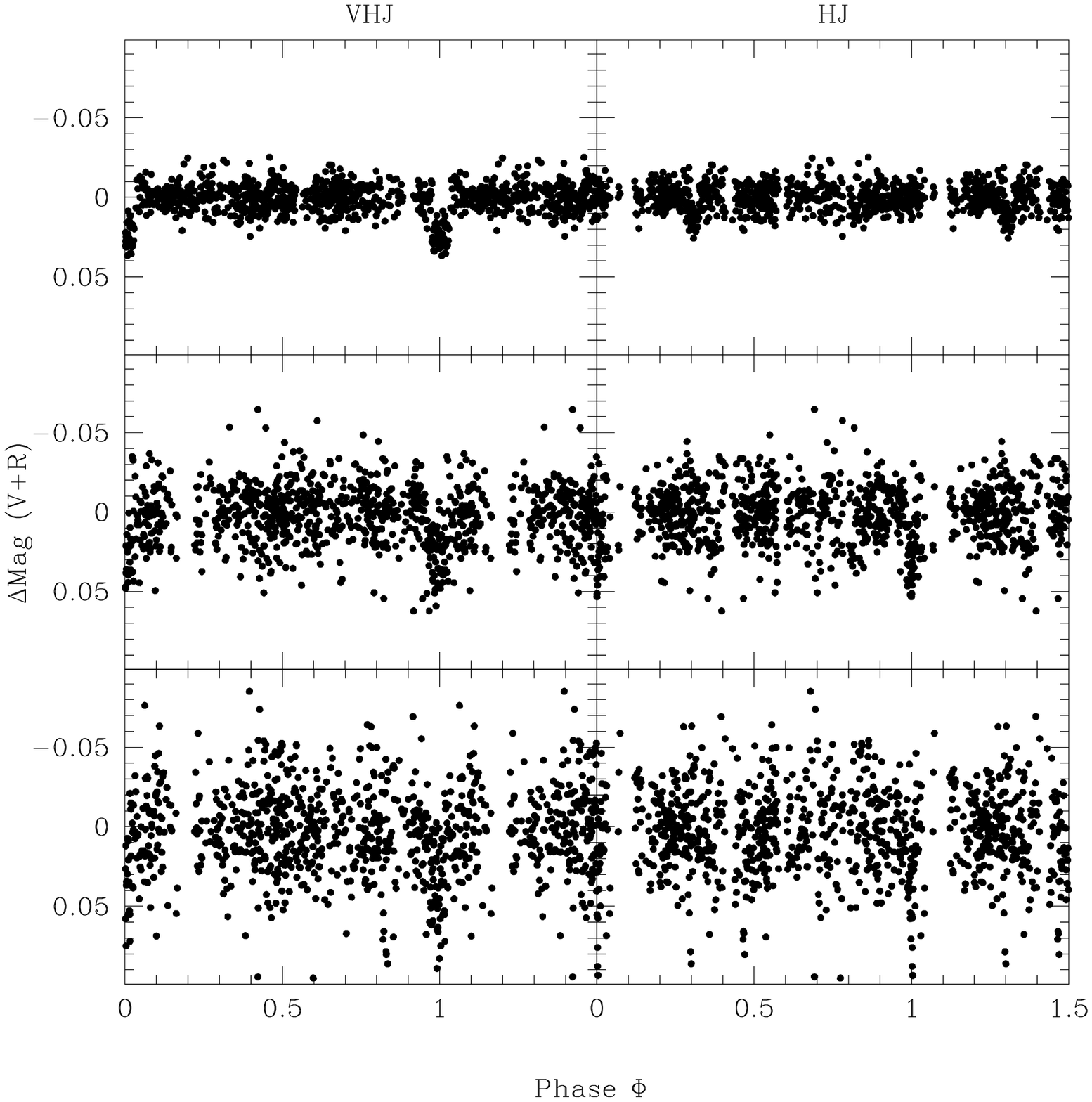]{Examples of the visibility of artificial transits producing output significances at the lowest end of detection significance (SDE $\sim$6 $\sigma$ with BLS, 11 $S(P_{\rm{mod}},\tau_{\rm{shift}})$ with WS). Left panels show a transit for a planet with an orbital period (P$_{\rm{orb}}$) of 1.5d (typical of a VHJ) with the right panels having P$_{\rm{orb}}=$3.3d (HJ). From top to bottom are transits for a 1.0R$_{\rm{Jup}}$ planet superimposed on a 1$\%$ photometry (foreground) star, a 1.5R$_{\rm{Jup}}$ planet with a 2$\%$ star and a 1.5R$_{\rm{Jup}}$ with a 4$\%$ star. The transits have been produced with parameters appropriate for the radius of the host star (Table.\space1).\label{fakesplot}}

\figcaption[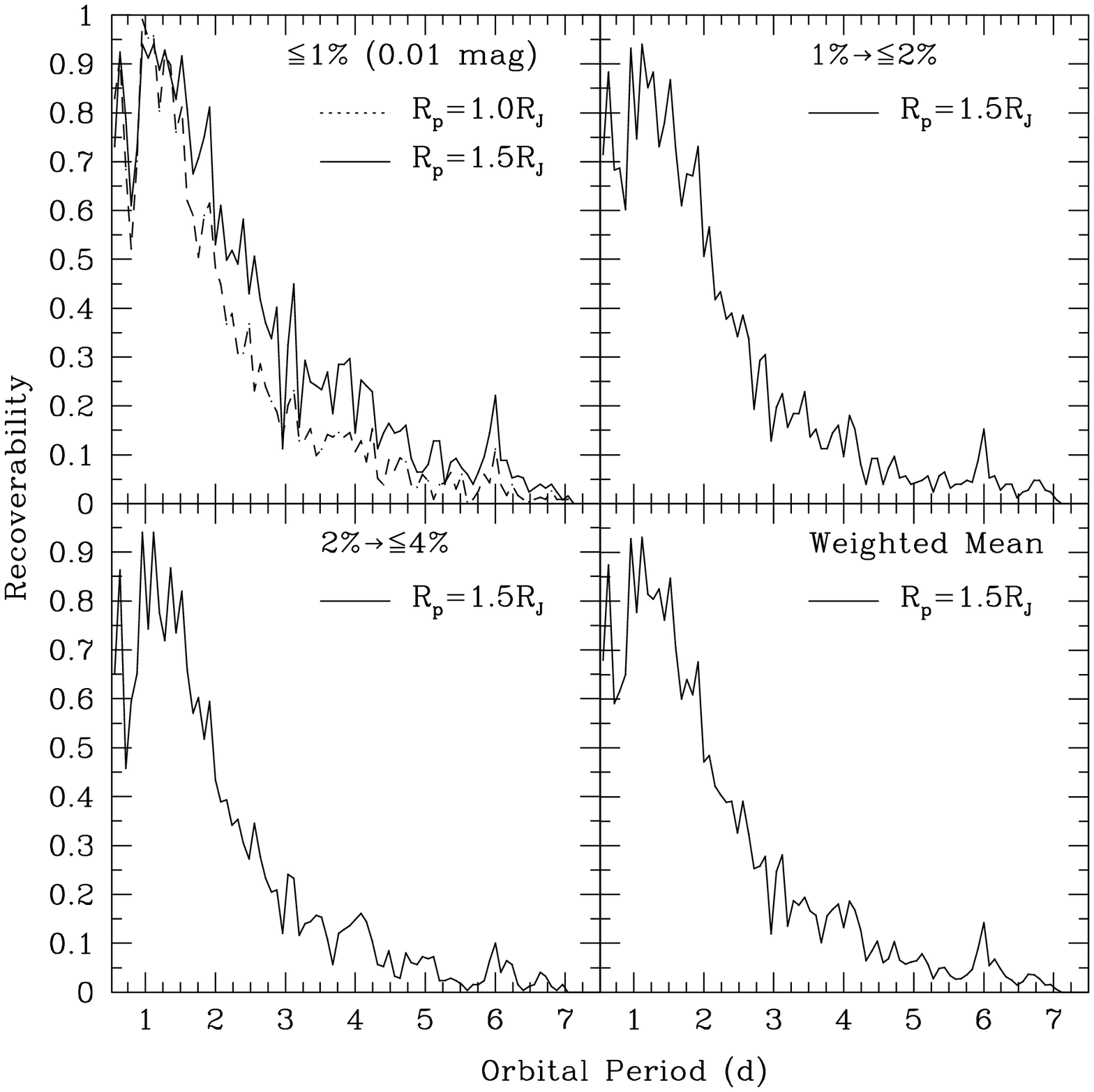]{The statistical transit recoverability for a range of planetary radii for stars with photometric precision $\le$0.01mag (top left) and 0.01$\rightarrow$0.02 mag (top right), 0.02$\rightarrow$0.04 mag (bottom left) and weighted mean (bottom right) plotted as the fraction of artificial transits recovered as a function of orbital period in days. The top left panel corresponds to the brightest stars, hence Galactic disk contamination. The top right panel corresponds to stars on the upper 0.8 magnitudes of cluster main sequence. The drop in recoverability near 1d is due to diurnal effects. These window functions indicate the necessity of transit surveys to contain many data points over a long observational period in order to be statistically sensitive to longer period planets. The transit recoverability for stars towards the faintest end of our search range (bottom left), typical of stars from V$\sim$18.2$\rightarrow$19.0. The bottom right panel shows the weighted mean transit recoverability from all Monte Carlo simulations across all stars, calculated accounting for the relative numbers of cluster main sequence stars in our search range. This recoverability was used in calculating the expected number of planets that should be detectable in our dataset.\label{mchist}}

\figcaption[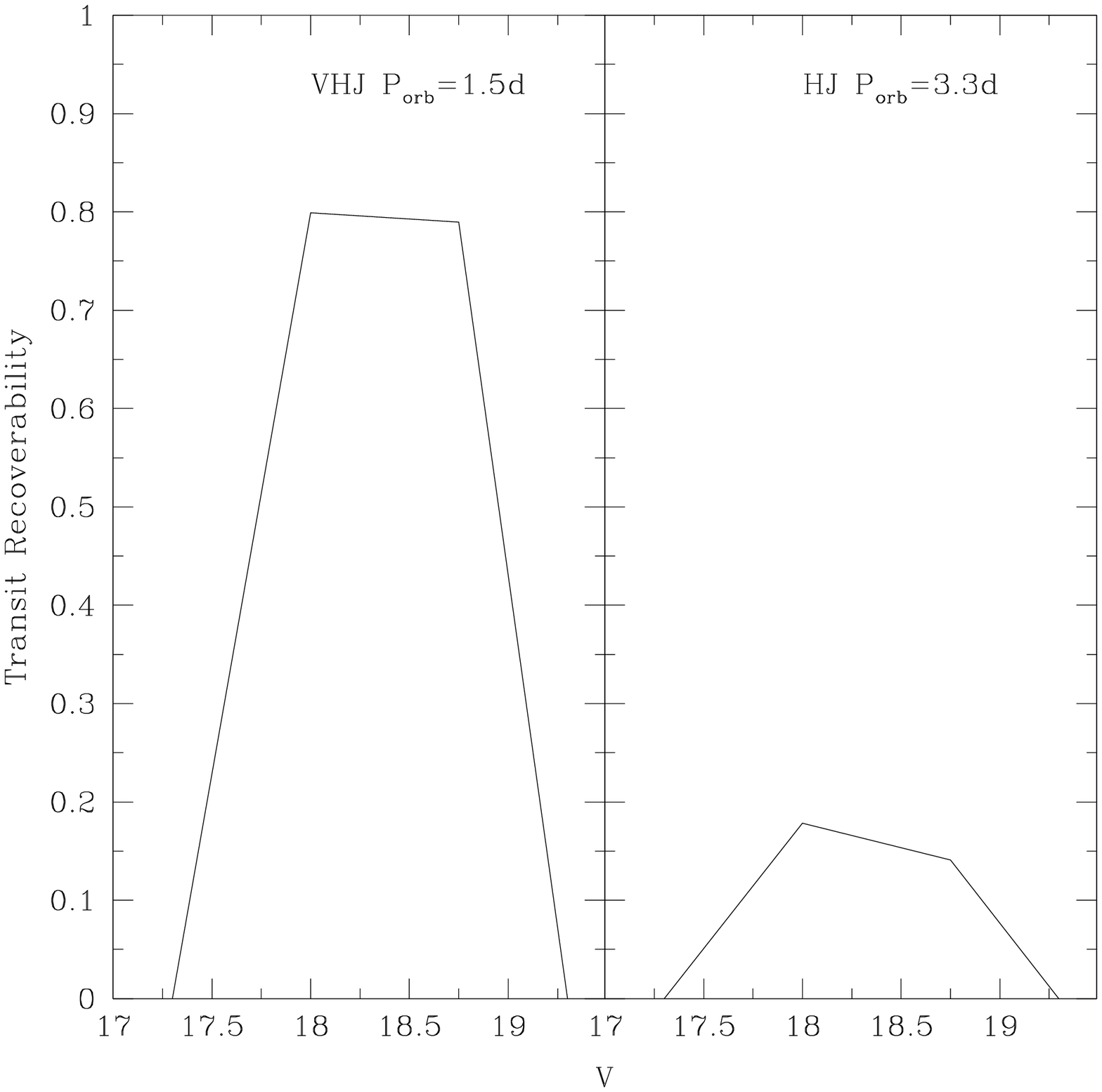]{The recoverability fraction of planetary transits in our dataset for R$=$1.5R$_{\rm{Jup}}$ planets as a function of V magnitude along the cluster main sequence. The limits to the recoverability are produced by the larger sizes of cluster stars above the cluster main sequence (V$\le$17.5) and the increasing photometric uncertainty (V$\ge$19.5).\label{magnumplot}}

\figcaption[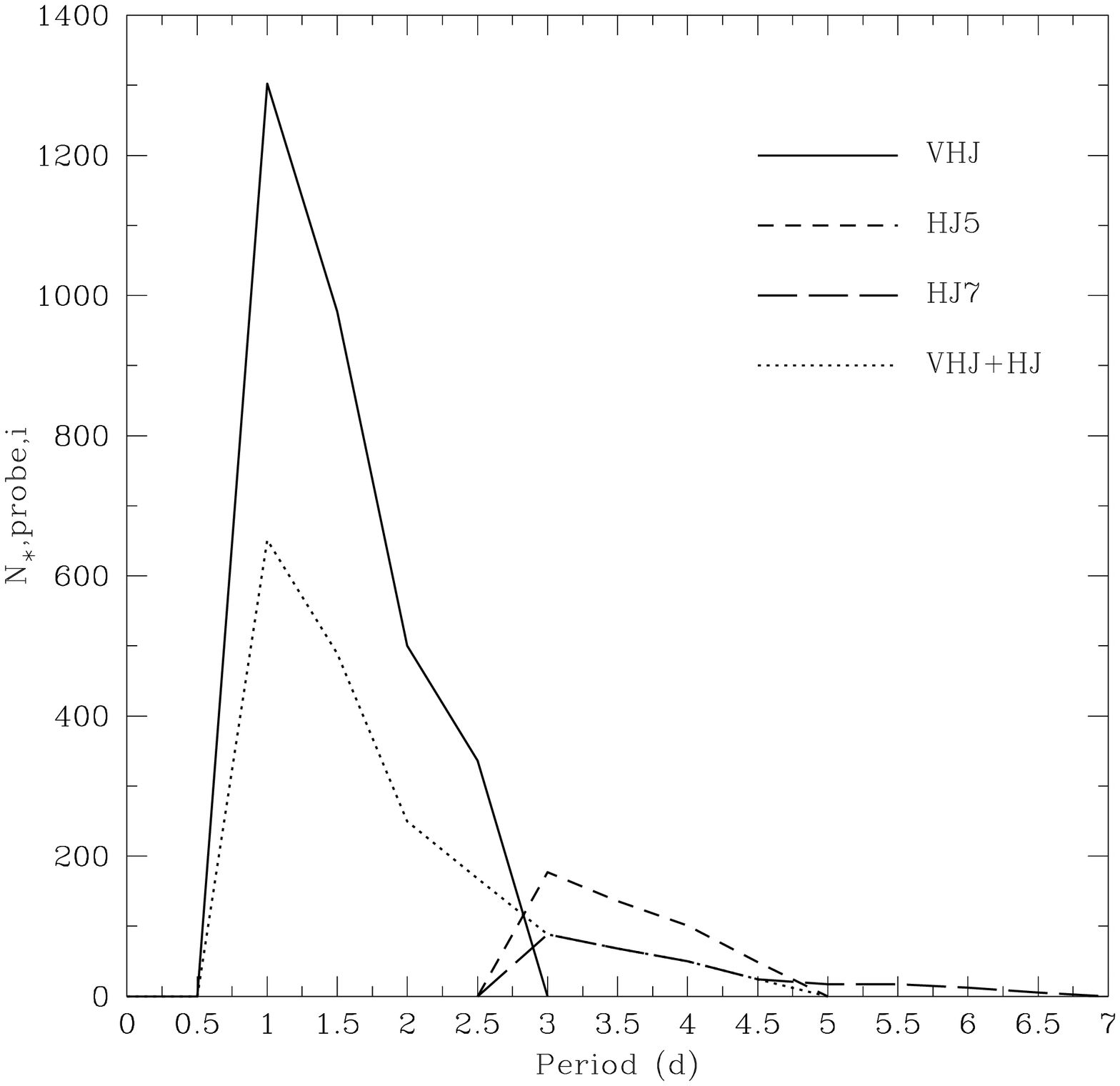]{The total number of stars probed for full sensitivity to transiting large radius planets in the cluster, distributed in orbital period for each of our four period distribution hypotheses. The solid line denotes the number of stars for VHJ planets distributed equally over 1.0$<P<$3.0d, the short-dashed line shows the star numbers for HJ planets of 3.0$<P<$5.0d, with the long-dashed line for HJ planets of 3.0$<P<$7.0d and finally the dotted line for VHJ and HJ distributed equally over 1.0$<P<$5.0d.\label{expnums}}

\clearpage

\begin{figure}
\plotone{f1.eps}
\centerline{f1.eps}
\end{figure}
\clearpage

\begin{figure}
\plotone{f2.eps}
\centerline{f2.eps}
\end{figure}
\clearpage

\begin{figure}
\plotone{f3.eps}
\centerline{f3.eps}
\end{figure}
\clearpage

\begin{figure}
\plotone{f4.eps}
\centerline{f4.eps}
\end{figure}
\clearpage

\begin{figure}
\plotone{f5.eps}
\centerline{f5.eps}
\end{figure}
\clearpage

\begin{figure}
\plotone{f6.eps}
\centerline{f6.eps}
\end{figure}
\clearpage

\begin{figure}
\plotone{f7.eps}
\centerline{f7.eps}
\end{figure}
\clearpage

\begin{figure}
\plotone{f8.eps}
\centerline{f8.eps}
\end{figure}


\begin{thebibliography}{}
\bibitem[Aigrain \& Pont(2007)]{Ai2007} Aigrain, S., \& Pont, 
F.\ 2007, \mnras, 378, 741
\bibitem[Aigrain et al.(2007)]{A2007} Aigrain, S., Hodgkin, 
S., Irwin, J., Hebb, L., Irwin, M., Favata, F., Moraux, E., \& Pont, F.\ 
2007, \mnras, 375, 29 
\bibitem[Alard \& Lupton(1998)]{AL98} Alard, C.~\& Lupton, 
R.~H.\ 1998, \apj, 503, 325 
\bibitem[Alonso et al.(2004)]{A2004} Alonso, R., et al.\ 
2004, \apjl, 613, L153 
\bibitem[Bakos et al.(2007c)]{Bakos2007c} Bakos, G.~A., et al.\ 
2007, ArXiv e-prints, 710, arXiv:0710.1841 
\bibitem[Bakos et al.(2007b)]{Bakos2007} Bakos, G.~A., et al.\ 
2007, ArXiv e-prints, 705, arXiv:0705.0126 
\bibitem[Bakos et al.(2007a)]{Bakos2006} Bakos, G.~{\'A}., et al.\ 
2007, \apj, 656, 552 
\bibitem[Barbieri et al.(2007)]{Barb2007} Barbieri, M., et al.\ 
2007, ArXiv e-prints, 710, arXiv:0710.0898 
\bibitem[Bekki \& Freeman(2003)]{BF2003} Bekki, K., \& 
Freeman, K.~C.\ 2003, \mnras, 346, L11 
\bibitem[Bekki \& Norris(2006)]{BN2005} Bekki, K., \& Norris, 
J.~E.\ 2006, \apjl, 637, L109 
\bibitem[Bouchy et al.(2004)]{B2004} Bouchy, F., Pont, F., 
Santos, N.~C., Melo, C., Mayor, M., Queloz, D., \& Udry, S.\ 2004, \aap, 
421, L13 
\bibitem[Bouchy et al.(2005)]{B2005} Bouchy, F., et al.\ 
2005, \aap, 444, L15 
\bibitem[Bruntt et al.(2003)]{Bruntt03} Bruntt, H., Grundahl, 
F., Tingley, B., Frandsen, S., Stetson, P.~B., \& Thomsen, B.\ 2003, \aap, 
410, 323 
\bibitem[Burke et al.(2007)]{Bu2007} Burke, C.~J., et al.\ 2007, ArXiv e-prints, 705, arXiv:0705.0003 
\bibitem[Burke et al.(2006)]{Bu2005} Burke, C.~J., Gaudi, 
B.~S., DePoy, D.~L., \& Pogge, R.~W.\ 2006, \aj, 132, 210 
\bibitem[Charbonneau et al.(2000)]{C2000} Charbonneau, D., 
Brown, T.~M., Latham, D.~W., \& Mayor, M.\ 2000, \apjl, 529, L45 
\bibitem[Coleman(2004)]{C2004} Coleman, M.~G. \ 2004, PhD.Thesis ``Tidal Structure in Galactic Satellites'' Australian National University
\bibitem[Cameron et al.(2007)]{CC2006} Cameron, A.~C., et al.\ 
2007, \mnras, 375, 951 
\bibitem[Dickens \& Woolley(1967)]{DW1967} Dickens, R.~J., \& 
Woolley, R.~v.~d.~R.\ 1967, Royal Greenwich Observatory Bulletin, 128, 255 
\bibitem[Dinescu et al.(1999)]{D1999} Dinescu, D.~I., Girard, 
T.~M., \& van Altena, W.~F.\ 1999, \aj, 117, 1792 
\bibitem[Fischer \& Valenti(2005)]{FV2005} Fischer, D.~A., \& 
Valenti, J.\ 2005, \apj, 622, 1102 
\bibitem[Fregeau et al.(2006)]{Freg2006} Fregeau, J.~M., 
Chatterjee, S., \& Rasio, F.~A.\ 2006, \apj, 640, 1086 
\bibitem[Gilliland et al.(2000)]{G2000} Gilliland, R.~L., et 
al.\ 2000, \apjl, 545, L47 
\bibitem[Gillon et al.(2007)]{G2007} Gillon, M., et al.\ 
2007, \aap, 472, L13 
\bibitem[Gonzalez(1997)]{G1997} Gonzalez, G.\ 1997, \mnras, 
285, 403
\bibitem[Gould et al.(2006)]{Gouldetal2006} Gould, A., Dorsher, S., 
Gaudi, B.~S., \& Udalski, A.\ 2006, Acta Astronomica, 56, 1  
\bibitem[Harris(1996)]{Harris96} Harris, W.~E.\ 1996, \aj, 112, 
1487 
\bibitem[Hartman et al.(2004)]{H2004} Hartman, J.~D., Bakos, 
G., Stanek, K.~Z., \& Noyes, R.~W.\ 2004, \aj, 128, 1761 
\bibitem[Henry et al.(2000)]{He2000} Henry, G.~W., Marcy, 
G.~W., Butler, R.~P., \& Vogt, S.~S.\ 2000, \apjl, 529, L41 
\bibitem[Hidas et al.(2005)]{H2005} Hidas, M.~G., et al.\ 
2005, \mnras, 360, 703 
\bibitem[Hood et al.(2005)]{Ho2005} Hood, B., et al.\ 2005, 
\mnras, 360, 791 
\bibitem[Horne(2003)]{H2003} Horne, K.\ 2003, ASP 
Conf.~Ser.~294: Scientific Frontiers in Research on Extrasolar Planets, 
294, 361 
\bibitem[Ideta \& Makino(2004)]{IM2004} Ideta, M., \& Makino, 
J.\ 2004, \apjl, 616, L107 
\bibitem[Janes(1996)]{J1996} Janes, K.\ 1996, \jgr, 101, 
14853 
\bibitem[Jenkins, Doyle \& Cullers(1996)]{Jenk96} Jenkins, 
J.~M., Doyle, L.~R., \& Cullers, D.~K.\ 1996, Icarus, 119, 244 
\bibitem[Kane et al.(2005)]{Ka2005} Kane, S.~R., Collier 
Cameron, A., Horne, K., James, D., Lister, T.~A., Pollacco, D.~L., Street, 
R.~A., \& Tsapras, Y.\ 2005, \mnras, 364, 1091 
\bibitem[Konacki et al.(2003)]{K2003} Konacki, M., Torres, 
G., Jha, S., \& Sasselov, D.~D.\ 2003, \nat, 421, 507 
\bibitem[Konacki et al.(2005)]{K2005} Konacki, M., Torres, 
G., Sasselov, D.~D., \& Jha, S.\ 2005, \apj, 624, 372 
\bibitem[Kovacs et al.(2007)]{Kov2007} Kovacs, G., et al.\ 
2007, ArXiv e-prints, 710, arXiv:0710.0602 
\bibitem[Kov{\'a}cs et al.(2002)]{K2002} Kov{\'a}cs, G., 
Zucker, S., \& Mazeh, T.\ 2002, \aap, 391, 369 
\bibitem[Laughlin(2000)]{L2000} Laughlin, G.\ 2000, \apj, 
545, 1064
\bibitem[Lee et al.(1999)]{Lee1999} Lee, Y.-W., Joo, J.-M., 
Sohn, Y.-J., Rey, S.-C., Lee, H.-C., \& Walker, A.~R.\ 1999, \nat, 402, 55 
\bibitem[Mandushev et al.(2007)]{Mand07} Mandushev, G., et 
al.\ 2007, ArXiv e-prints, 708, arXiv:0708.0834 
\bibitem[Marcy et al.(2005)]{M2005} Marcy, G., Butler, R.~P., 
Fischer, D., Vogt, S., Wright, J.~T., Tinney, C.~G., \& Jones, H.~R.~A.\ 
2005, Progress of Theoretical Physics Supplement, 158, 24 
\bibitem[McCullough et al.(2006)]{MCC2006} McCullough, P.~R., 
et al.\ 2006, \apj, 648, 1228 
\bibitem[Meylan et al.(1995)]{Meylan1995} Meylan, G., Mayor, M., 
Duquennoy, A., \& Dubath, P.\ 1995, \aap, 303, 761 
\bibitem[Meylan \& Mayor(1986)]{MM86} Meylan, G., \& Mayor, 
M.\ 1986, \aap, 166, 122 
\bibitem[Mochejska et al.(2003)]{M2003} Mochejska, B.~J., 
Stanek, K.~Z., Sasselov, D.~D., \& Szentgyorgyi, A.~H.\ 2003, IAU 
Symposium, 219,  
\bibitem[Mochejska et al.(2005)]{Mo2005} Mochejska, B.~J., et 
al.\ 2005, \aj, 129, 2856 
\bibitem[Mochejska et al.(2006)]{Moch2006} Mochejska, B.~J., et 
al.\ 2006, \aj, 131, 1090 
\bibitem[Norris \& Bessell(1975)]{NB1975} Norris, J., \& 
Bessell, M.~S.\ 1975, \apjl, 201, L75 
\bibitem[Norris(2004)]{N2004} Norris, J.~E.\ 2004, \apjl, 
612, L25 
\bibitem[Noyes et al.(2007)]{N2007} Noyes, R.~W., et al.\ 
2007, ArXiv e-prints, 710, arXiv:0710.2894 
\bibitem[O'Donovan et al.(2007)]{OD2007} O'Donovan, F.~T., et 
al.\ 2007, \apjl, 663, L37 
\bibitem[O'Donovan et al.(2006)]{OD2006} O'Donovan, F.~T., et al.\ 2006, \apjl, 651, L61 
\bibitem[Pancino et al.(2000)]{P2000} Pancino, E., Ferraro, 
F.~R., Bellazzini, M., Piotto, G., \& Zoccali, M.\ 2000, \apjl, 534, L83 
\bibitem[Pepper \& Gaudi(2005)]{PG2005} Pepper, J., \& Gaudi, 
B.~S.\ 2005, \apj, 631, 581 
\bibitem[Pepper \& Gaudi(2006)]{PG2006} Pepper, J., \& Gaudi, 
B.~S.\ 2006, Acta Astronomica, 56, 183 
\bibitem[Pepper et al.(2007)]{Pepper07} Pepper, J., Stanek, 
K.~Z., Pogge, R.~W., Latham, D.~W., DePoy, D.~L., Siverd, R., Poindexter, 
S., \& Sivakoff, G.~R.\ 2007, ArXiv e-prints, 709, arXiv:0709.2728 
\bibitem[Pont et al.(2004)]{P2004} Pont, F., Bouchy, F., 
Queloz, D., Santos, N.~C., Melo, C., Mayor, M., \& Udry, S.\ 2004, \aap, 
426, L15 
\bibitem[Robin et al.(2003)]{Rob2003} Robin, A.~C., Reyl{\'e}, 
C., Derri{\`e}re, S., \& Picaud, S.\ 2003, \aap, 409, 523 
\bibitem[Sahu et al.(2006)]{Sahu2006} Sahu, K.~C., et al.\ 2006, 
\nat, 443, 534 
\bibitem[Sato et al.(2005)]{S2005} Sato, B., et al.\ 2005, 
\apj, 633, 465 
\bibitem[Santos et al.(2001)]{S2001} Santos, N.~C., 
Israelian, G., \& Mayor, M.\ 2001, \aap, 373, 1019 
\bibitem[Sigurdsson(1992)]{Sig1992} Sigurdsson, S.\ 1992, 
\apjl, 399, L95 
\bibitem[Sigurdsson et al.(2003)]{Sig2003} Sigurdsson, S., 
Richer, H.~B., Hansen, B.~M., Stairs, I.~H., \& Thorsett, S.~E.\ 2003, 
Science, 301, 193 
\bibitem[Sollima et al.(2005)]{Sol2005} Sollima, A., Ferraro, 
F.~R., Pancino, E., \& Bellazzini, M.\ 2005, \mnras, 357, 265 
\bibitem[Street et al.(2003)]{S2003} Street, R.~A., et al.\ 
2003, ASP Conf.~Ser.~294: Scientific Frontiers in Research on Extrasolar 
Planets, 294, 401 
\bibitem[Tamuz et al.(2005)]{T2005} Tamuz, O., Mazeh, T., \& 
Zucker, S.\ 2005, \mnras, 356, 1466 
\bibitem[Tingley(2003a)]{Ting03a} Tingley, B.\ 2003a, \aap, 403, 
329 
\bibitem[Tingley(2003b)]{Ting03b} Tingley, B.\ 2003b, \aap, 408, 
L5 
\bibitem[Udalski et al.(2002)]{U2002} Udalski, A., et al.\ 
2002, Acta Astronomica, 52, 1 
\bibitem[Udalski et al.(2004)]{U2004} Udalski, A., Szymanski, 
M.~K., Kubiak, M., Pietrzynski, G., Soszynski, I., Zebrun, K., Szewczyk, 
O., \& Wyrzykowski, L.\ 2004, Acta Astronomica, 54, 313 
\bibitem[van de Ven et al.(2006)]{VDV2006} van de Ven, G., van 
den Bosch, R.~C.~E., Verolme, E.~K., \& de Zeeuw, P.~T.\ 2006, \aap, 445, 
513
\bibitem[von Braun et al.(2005)]{V2005} von Braun, K., Lee, 
B.~L., Seager, S., Yee, H.~K.~C., Mall{\'e}n-Ornelas, G., \& Gladders, 
M.~D.\ 2005, \pasp, 117, 141 
\bibitem[Weldrake et al.(2007)]{W2007a} Weldrake, D.~T.~F., 
Bayliss, D.~D.~R., Sackett, P.~D., Bessell, M., \& Tingley, B.\ 2007, 
Transiting Extrapolar Planets Workshop, 366, 90 
\bibitem[Weldrake et al.(2007b)]{W2006} Weldrake, D.~T.~F., 
Sackett, P.~D., \& Bridges, T.~J.\ 2007, \aj, 133, 1447 
\bibitem[Weldrake et al.(2005)]{W2005} Weldrake, D.~T.~F., 
Sackett, P.~D., Bridges, T.~J., \& Freeman, K.~C.\ 2005, \apj, 620, 1043 
\bibitem[Weldrake \& Sackett(2005)]{WS2005} Weldrake, 
D.~T.~F., \& Sackett, P.~D.\ 2005, \apj, 620, 1033 
\bibitem[Wozniak(2000)]{Woz2000} Wozniak, P.~R.\ 2000, Acta 
Astronomica, 50, 421 
\bibitem[Yi et al.(2003)]{Y2003} Yi, S.~K., Kim, Y.-C., \& 
Demarque, P.\ 2003, \apjs, 144, 259
\end{thebibliography}
\end{document}